\documentclass[prb,aps,twocolumn,showpacs,superscriptaddress,10pt]{revtex4-2}
\usepackage{amsfonts,amsmath,bm,multirow}
\usepackage[OT4]{fontenc}

\usepackage{graphicx}
\usepackage{xcolor}
\usepackage{epstopdf}
\usepackage{dsfont}
\usepackage{physics}
\usepackage{hyperref}
\usepackage{lipsum}
\usepackage{dcolumn}

\usepackage{natbib}
\usepackage{hf-tikz}
\usepackage[utf8]{inputenc}

\newcommand{\kp}{\mbox{$\bm{k}\!\vdot\!\bm{p}\ $}}
\newcommand{\rr}{\bm{r}}

\definecolor{bred}{HTML}{e31a1c}
\definecolor{bgreen}{HTML}{33a02c}
\definecolor{bblue}{HTML}{1f78b4}

\definecolor{armygreen}{rgb}{0.29, 0.33, 0.13}

\newcolumntype{L}{D{.}{.}{3,4}}

\begin{document}

\title {Resonant and Anti-resonant Exciton-Phonon Coupling in Quantum Dot Molecules}

\author{Michelle Lienhart$^{**}$}
\email{michelle.lienhart@tum.de}
\affiliation{Walter Schottky Institut, School of Natural Sciences, and MCQST, Technische Universität München, 85748 Garching, Germany}

\author{Krzysztof Gawarecki$^{**}$}
\email{krzysztof.gawarecki@pwr.edu.pl}
\affiliation{Institute of Theoretical Physics, Wroc\l{}aw University of Science and Technology, Wroc\l{}aw 50-370, Poland}

\author{Markus St\"ocker}
\affiliation{Walter Schottky Institut, School of Natural Sciences, and MCQST, Technische Universität München, 85748 Garching, Germany}

\author{Frederik Bopp}
\affiliation{Walter Schottky Institut, School of Natural Sciences, and MCQST, Technische Universität München, 85748 Garching, Germany}

\author{Charlotte Cullip}
\affiliation{Walter Schottky Institut, School of Natural Sciences, and MCQST, Technische Universität München, 85748 Garching, Germany}

\author{Nadeem Akhlaq}
\affiliation{Walter Schottky Institut, School of Natural Sciences, and MCQST, Technische Universität München, 85748 Garching, Germany}

\author{Christopher Thalacker}
\affiliation{Walter Schottky Institut, School of Natural Sciences, and MCQST, Technische Universität München, 85748 Garching, Germany}

\author{Johannes Schall}
\affiliation{Institut f\"ur Physik und Astronomie, Technische Universit\"at Berlin, Hardenbergstra\ss e 36, 10623 Berlin, Germany}

\author{Sven Rodt}
\affiliation{Institut f\"ur Physik und Astronomie, Technische Universit\"at Berlin, Hardenbergstra\ss e 36, 10623 Berlin, Germany}

\author{Arne Ludwig}
\affiliation{Lehrstuhl f\"ur Angewandte Festk\"orperphysik, Ruhr-Universit\"at Bochum, Universit\"atsstra\ss e 150, 44801 Bochum, Germany}

\author{Dirk Reuter}
\affiliation{Paderborn University, Department of Physics, 33098 Paderborn, Germany}

\author{Stephan Reitzenstein}
\affiliation{Institut f\"ur Physik und Astronomie, Technische Universit\"at Berlin, Hardenbergstra\ss e 36, 10623 Berlin, Germany}

\author{Kai M\"uller}
\affiliation{Walter Schottky Institut, School of Computation, Information and Technology, and MCQST, Technische Universität München, 85748 Garching, Germany}

\author{Pawe\l{} Machnikowski}
\affiliation{Institute of Theoretical Physics, Wroc\l{}aw University of Science and Technology, Wroc\l{}aw 50-370, Poland}

\author{Jonathan J. Finley}
\email{jj.finley@tum.de \newline\newline$^{**}$ These authors contributed equally to this work.}
\affiliation{Walter Schottky Institut, School of Natural Sciences, and MCQST, Technische Universität München, 85748 Garching, Germany}

\begin{abstract}
Optically active quantum dot molecules (QDMs) can host multi-spin quantum states with the potential for the deterministic generation of photonic graph states with tailored entanglement structures. Their usefulness for the generation of such non-classical states of light is determined by orbital and spin decoherence mechanisms, particularly phonon-mediated processes dominant at energy scales up to a few millielectronvolts.
Here, we directly measure the spectral function of orbital phonon relaxation in a QDM and benchmark our findings against microscopic $\kp$ theory.  Our results reveal phonon-mediated relaxation rates exhibiting pronounced resonances and anti-resonances, with rates ranging from several ten ns$^{-1}$ to tens of $\mu$s$^{-1}$.
Comparison with a kinetic model reveals the voltage (energy) dependent phonon coupling strength and fully explains the interplay between phonon-assisted relaxation and radiative recombination.
These anti-resonances can be leveraged to increase the lifetime of energetically unfavorable charge configurations needed for realizing efficient spin-photon interfaces and multi-dimensional cluster states.
\end{abstract}

\maketitle

\section{Introduction}
\label{sec:intr}
Distributed photon based quantum technologies using solid-state systems typically involve links between flying qubits in the form of photons and static qubits such as spins in solid state materials. To enable spin qubits, the spins should have a long coherence time, a strong light-matter interaction, and a tunable resonance to enable applications in photonic quantum information technology \cite{Awschalom2018}.
Optically active semiconductor quantum dots (QDs) fulfill these requirements by: possessing robust polarization selection rules that enable spin-to-optical polarization mapping \cite{Bayer2002, Gao2012, Stockill2017}, exhibiting dominant zero-phonon line emission at low temperatures \cite{Favero2003AcousticPS}, and demonstrating nearly Fourier-limited optical linewidths \cite{Kuhlmann2015}.
These characteristics make QDs excellent candidates for spin-photon interfaces, particularly for the deterministic generation of one-dimensional photonic graph states \cite{lindnerProposalPulsedOnDemand2009, Cogan2023}.
Quantum dot molecules (QDMs), consisting of two vertically stacked and tunnel-coupled QDs, offer significant advantages over single QDs. The singlet-triplet ($S$-$T_0$) configuration of two spins makes them insensitive to magnetic and electrical noise at specific applied external electric fields, yielding coherence times $T_2^{*}$ more than an order of magnitude longer than single spins \cite{tranEnhancedSpinCoherence2022, Doty2010}.
QDMs also feature electrically tunable orbital structure \cite{Ardelt2016} and transition dipole moments \cite{Bopp2023a}, and they can be optically charged \cite{boppQuantumDotMolecule2022} to prepare two-spin states needed for creating one- and two-dimensional entanglement structures required in measurement based quantum information processing \cite{economouOpticallyGenerated2Dimensional2010,lindnerProposalPulsedOnDemand2009,vezvaee2022}.
To facilitate these applications, QDMs must maintain
high cluster state creation fidelities, which are fundamentally limited by spin dephasing mechanisms, particularly phonon-induced decoherence.

The main processes that limit the lifetime of excited orbital states in QDMs and, consequently, affect the orbital and spin dephasing are exciton recombination due to spontaneous emission of photons and phonon-induced decoherence in the form of relaxation \cite{nakaokaDirectObservationAcoustic2006} or pure dephasing \cite{Borri2001}. In most of the published works, these two classes of processes were treated as additive \cite{kawa2022} ``one-way'' impact of phonon processes on optical emission \cite{Wiercinski2023} or optical control \cite{Ramsay2010} was taken into account, or optical signatures of phonon-related effects were discussed \cite{Vagov2004}. This can usually be justified by the large difference between the picosecond time scales over which phonon-induced processes typically occur and nanosecond radiative lifetimes. 
The disparity between timescales does not necessarily hold in QDMs, since the interdot tunnel coupling and the field-tunable energy between excitons of the vertically stacked QDs allow one to bring the radiative decay rates  on the scale of interdot phonon-assisted relaxation \cite{nakaokaDirectObservationAcoustic2006}. In this regime, the inter-band transitions couple to the low energy tails of phonon spectral densities. The carrier-phonon coupling depends sensitively on the geometry of the charge carrier wave functions \cite{Knorr2006}.
Hence, the resulting relaxation dynamics depends on the morphology of the system. In fact, the details of the coupling can in some cases be inferred from the observed phonon-induced carrier dynamics manifested in the optical response \cite{Wijesundara2011}.

Here, we combine experiment and theory to explore signatures of combined carrier-phonon-photon kinetics in which both dissipative processes have comparable rates.  We measure the lifetime of the neutral exciton ($X^0$) branches in an electrically tunable QDM and observe non-monotonic behaviour of the exciton lifetime as a function of the applied voltage.
Our kinetic model incorperates quantitative modeling of carrier states and phonon spectral densities using \kp theory based on precise mapping of the QDM morphology and composition profile \cite{Gawarecki2018a}.
This allows us to relate this feature to the interplay between phonon-assisted relaxation and radiative recombination in a regime where relaxation is slowed down by tuning the energy separation between the two lowest $X^0$ eigenenergies in the QDM  to the oscillating tail of the phonon spectral density.
The phonon-mediated relaxation rates we observe exhibit resonances where relaxation is maximal and anti-resonances (relaxation minima) where the system is protected from phonon-mediated relaxation.
More than ten years ago, theoretical studies have predicted these resonances multiple times more than ten years ago \cite{Climente2006, Gawarecki2010, Gawarecki2012, Grodecka-Grad2010}. Experiments as well have observed phonon-mediated relaxation effects in QDMs in the past \cite{Wijesundara2011,Mueller2012}, however, theory and experiment have not yet been fully reconciled.
These anti-resonances, where the phonon-mediated coupling of orbital and spin states are strongly suppressed, can be used to circumvent phonon-mediated relaxation and decoherence processes. This ability is likely to have significant advantages for the use of entangled spin states for the generation of multi-dimensional cluster states for quantum communication and measurement based quantum computation\cite{vezvaee2022}.
	
The paper is organized as follows: In Sec.~\ref{sec:exp}, we describe the sample design, experiments, and results of the voltage dependent time resolved measurements. We also present the effective model utilized to extract the exciton kinetics. In Sec.~\ref{sec:model}, we define the theoretical model, which is used to quantitatively interpret the experimental results. Finally, Sec.~\ref{sec:conclusions} contains the conclusions and compares the experiment with the theory.

\section{Experiment}
\label{sec:exp}

In this section we describe the experimental methods, present the measurement results, and define a simple effective model that is used to interpret the measurements and to extract key parameters. 

\subsection{Sample structure and methodology}
\label{subsec:sample_methods}

\begin{figure}[tb!!]
    \begin{center}
        \includegraphics[width=0.48\textwidth]{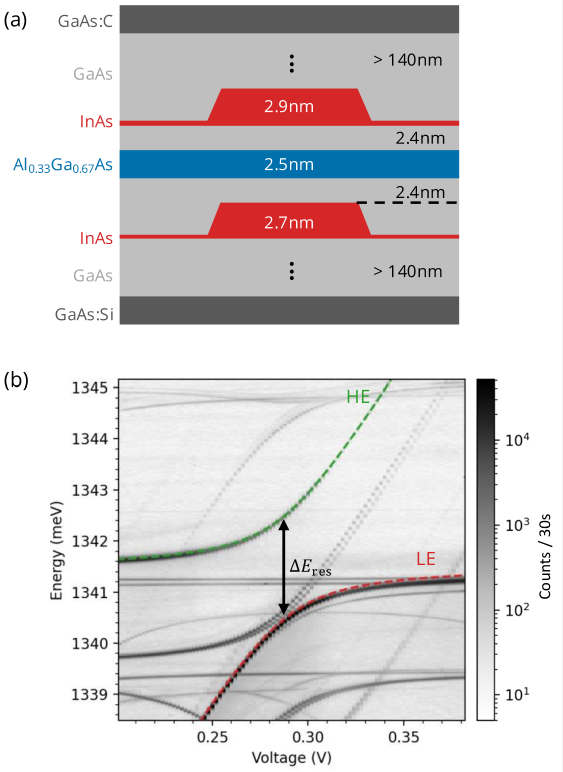}
    \end{center}
    \caption{\label{fig:sample} (a) Schematic structure showing the central part of the QDM sample embedded in a p-i-n diode. 
    (b) Voltage-dependent photoluminescence spectra of a QDM. The neutral exciton eigenenergies (HE in green and LE in red) were fitted by a two-state effective Hamiltonian (see Eq.\,\ref{eq:eff}). $\Delta E_{\mathrm{res}}$ is the energy splitting at the resonance condition at $0.288\,$V.
    }
\end{figure}

Figure \,\ref{fig:sample}(a) shows the (epitaxial) layer design of the QDM sample, consisting of two vertically stacked InGaAs QDs embedded in a GaAs matrix and grown using the In-flush method to precisely control their absolute height. The inter-dot separation was controlled during growth to be 10\,nm (wetting layer of both dots), resulting in electron tunnel-coupling between the dots forming the molecule \cite{Krenner2005}.
The QDM is embedded in the intrinsic region of a p-i-n diode allowing the application of a gate voltage $V$ to the sample to control the static electric field along the QDM growth axis. 
The Al$_{0.33}$Ga$_{0.67}$As layer between the two dots acts as a tunnel barrier, thereby reducing the tunnel coupling strength.
Further information on the growth and fabrication of the sample can be found in the Appendix\,\ref{app:sample} and in previous work in \cite{Bopp2023a} and \cite{BoppMagTun}.

We optically address the fundamental neutral exciton transitions ($X^0$) as their energy splitting is tuned to explore signatures of interplay between the phonon-induced and radiative kinetics.
In all measurements presented here, we make use of a two-step electrical and optical sequence: Firstly, we efficiently prepare the QDM in a zero-charge (empty QDM) ground-state by applying a large negative voltage of -3\,V. In a subsequent measurement step, we apply a static gate voltage \cite{boppQuantumDotMolecule2022} to tune the electronic orbital states in each of the two QDs relative to each other.
The top dot has a larger height, thereby ensuring that the hole of an excited exciton is always located in the top dot.
In this way, the spatially direct (electron in top dot) and indirect (electron in bottom dot) exciton energies can be tuned into resonance.

To perform luminescence lifetime measurements the $X^0$ is excited using a 5\,ps laser pulse tuned to 1362.1\,meV to be resonant to the p-shell in the upper dot.
After a few picoseconds, the system relaxes non-radiatively to the energetically lower s-type exciton states, which enables the detection of the $X^0$ s-shell emission and its energy and lifetime. See Appendix~\ref{app:setup} for further details.

\subsection{Experimental results}

The voltage-dependent photoluminescence spectra of the neutral exciton is presented in Fig.\,\ref{fig:sample}(b).
Two distinct energy branches are visible, which are hence forth referred to as the low-energy (LE) branch and the high-energy (HE) branch.  The weak parabolic voltage dependent shifts for LE and HE correspond to excitonic states having direct character (e and h in same dot) while strongly linearly shifting transitions are spatially indirect excitons (e in lower dot, upper dot), that exhibit a large linear Stark shift due to their significantly larger static dipole moment.
At $0.288$\,V the electron orbitals of both dots are in resonance and electron tunnel coupling results in a pronounced avoided crossing between the orbital states, shown by $\Delta E_\mathrm{res}$ in Fig.\,\ref{fig:sample}(b).
Tracing through the avoided crossing corresponds to a transition between the spatially direct and indirect exciton states. 
This reflects the well-known structure of the tunnel resonance, where the electron envelope function changes its localization from one dot to the other upon moving through the anticrossing~\cite{Krenner2005,Mueller2011,Ardelt2016,Bopp2023a}.
At the point of minimal energy splitting $\Delta E_\mathrm{res} = 1.988$\,meV, the electron envelope functions hybridize via coherent tunnel coupling to form bonding and antibonding superpositions of the uncoupled orbitals.

The exciton energies are fitted using a coupled two-state effective model including the DC Stark shift for inter-dot excitons~\cite{Krenner2005, Ardelt2016}.
The full description can be found in the Appendix \ref{app:PLVSimulation}.
The eigenstates of the effective model are shown in Fig.~\ref{fig:sample}(b) by the green and red dashed lines exhibiting very good agreement with experiment.
In the following analysis in this section (Sec.\,\ref{sec:exp}), the results of the effective model are used to calculate the radiative decay rates of the HE and LE branches, allowing us to visualize the phonon resonances as a function of the energy splitting between coupled exciton states.

\begin{figure}[tb!!]
    \begin{center}
        \includegraphics[width=.45\textwidth]{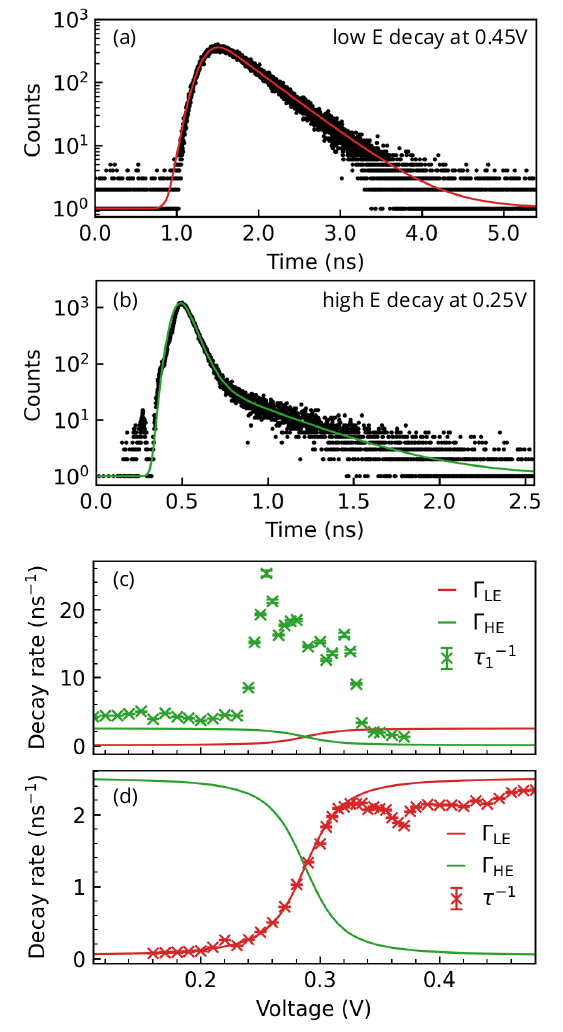}
    \end{center}
    \caption{\label{fig:decays_exp}
    The time dependence of the emission from the two lowest neutral exciton energy branches: (a) the exciton decay time from the lower energy branch at 0.45\,V and the fitting using an exponential function convolved with a Gaussian to account for the instrument response (IR); and (b) the results from the higher energy branch at 0.25\,V and the fitting by a sum of two exponential convolved with a Gaussian;
    (c) and (d) Gate voltage dependent decay times.
    The fast decay rates ${\tau_1}^{-1}$ of the higher energy branch are shown with green symbols in (c), the decay rates ${\tau}^{-1}$ of the lower energy branch with red symbols in (d).
    The red and green solid lines denote the radiative recombination rates $\Gamma_\mathrm{LE}$ and $\Gamma_\mathrm{HE}$ of LE and HE calculated from Eq.~\ref{eq:gamma_rad}, respectively. 
    }
\end{figure}

In order to probe the acoustic phonon couplings, time-resolved decay transients of the $X^0$ states were measured for various gate voltages around the avoided crossing.
In this region, the energy difference between HE and LE ranges from $\approx 2\,$meV to 10\,meV, making an acoustic phonon absorption or emission at $1.7$\,K possible.
Figure~\ref{fig:decays_exp}(a) and (b) show two representative decay curves: one recorded from the low-energy branch at 0.45\,V and the other from the high-energy branch at 0.25\,V.
The initial increase in the occupation for both cases comes from a nonradiative transitions from the higher orbital states (p-shell).
Fig.~\ref{fig:decays_exp}(a) shows a typical time resolved decay transient recorded from the low-energy branch (black).
The relaxation time is obtained by fitting with a mono-exponential function convolved with a Gaussian (red).
The exciton decay time of the high-energy branch (see Fig.~\ref{fig:decays_exp}(b) in black) exhibits two distinct slopes, corresponding to a fast \({\tau_1}^{-1}\) and a slow \({\tau_2}^{-1}\) relaxation time.
Their decay rates are obtained by fitting with a double-exponential decay convolved with a Gaussian function (green).
Details on the fitting functions and the convolution with an instrument response function can be found in Appendix~\ref{app:Decay_fitting}.

The measured decay times depend not only on the radiative lifetimes of the coupled orbital states, but also on other processes that depopulate the high-energy (HE) and low-energy (LE) states.  Whilst we will show below that these processes reflect the coupling to acoustic phonons, we first calculate the radiative recombination rates of the HE and LE branches.  
Therefore, we measure the radiative lifetime of the direct exciton by applying gate voltages far from any resonances and the avoided crossing, e.g. at $0.6$\,V, where the level splitting is approx. \(19\)\,meV).  Hereby, we suppress transitions mediated by acoustic phonons since the coupling matrix element vanishes for such large phonons as their wavelength is far below any critical dimension reflecting the envelope functions. 
As a result, the measured decay rate \(\tau^{-1} = 2.45\)\,ns\(^{-1}\) for the LE branch can be interpreted as the radiative recombination rate of the direct exciton, \(\Gamma_\mathrm{D} = \tau^{-1}\).  
The recombination rate for the indirect exciton, \(\Gamma_\mathrm{I} = 0.04\)\,ns\(^{-1}\), was estimated from the electric field dependent Rabi oscillation frequency, as reported in \cite{Bopp2023a}.
Using these determined values $\Gamma_\mathrm{D}$ and $\Gamma_\mathrm{I}$ far away from the avoided crossing, the radiative recombination rate can be calculated at arbitrary voltage in the spirit of the effective model, by weighting the recombination rates for the direct and indirect excitons according to the Hopfield coefficients of the coupled system,
\begin{equation}
    \label{eq:gamma_rad}
    \Gamma_n(\widetilde{F}) = \abs{c^{(\mathrm{D})}_n (\widetilde{F})}^2 \Gamma_{\mathrm{D}} + \abs{c^{(\mathrm{I})}_n (\widetilde{F})}^2 \Gamma_{\mathrm{I}},
\end{equation}
where $n \in \{ \mathrm{LE}, \mathrm{HE} \}$ labels the exciton energy branches, $\widetilde{F}$ is the gate voltage, and $c^{(\mathrm{D/I})}_n (\widetilde{F})$ are coefficients resulting from the diagonalization of the coupled two-state Hamiltonian (Eq.~\eqref{eq:eff} in Appendix~\ref{app:PLVSimulation}).
The results of this model are represented by the solid green (HE) and red (LE) lines in Fig.~\ref{fig:decays_exp}(c) and (d).
For gate voltages far below (above) $0.288$~V, the LE (HE) branch exhibits a strong indirect exciton character.
In this regime, the electron and hole reside in different QDs, leading to slow radiative recombination.
As the voltage increases (decreases), the overlap between the electron and hole wave functions increases, resulting in faster decay.
For gate voltages far above (below) $0.288$~V, the LE (HE) state is predominantly a direct exciton state.
The crossing of the solid lines marks the direct-indirect exciton transition, reflecting the hybridization of the electron wavefunction, as also illustrated in Fig.~\ref{fig:sample}(b).

The extracted decay rates of the high-energy (green crosses) and low-energy (red crosses) branches are shown in Fig.~\ref{fig:decays_exp}(c) and (d) for gate voltages ranging from $0.11$\,V to $0.48$\,V.
As described earlier, $\tau^{-1}$ for the low-energy branch is obtained by fitting a mono-exponential deacay.
The measured values of $\tau^{-1}$ closely follow the dependence predicted by the effective model for radiative decay, $\Gamma_\mathrm{LE}(\widetilde{F})$.
However, at approximately $0.36$\,V, a pronounced dip appears, corresponding to an HE-LE energy splitting of 4.8\,meV.  
The origin of this phonon-related feature is discussed in detail in Sec.~\ref{sec:results_theory}.

The high-energy branch is modeled using a double-exponential deacay, incorporating both \(\tau_1^{-1}\) (fast decay) and \(\tau_2^{-1}\) (slow decay) components. The lifetime \(\tau_1\) (fast decay) corresponds to the radiative decay of the HE state, further enhanced by phonon-assisted relaxation to the LE state.  
The second lifetime, \(\tau_2\) (slow decay), is associated with transitions from the LE state to the HE state via acoustic phonon absorption.  
The impact of this process is proportional to the occupation of the LE state, which itself depends on the radiative recombination rate \(\Gamma_\mathrm{LE}(\widetilde{F})\).
Consequently, the \(\tau_2^{-1}\) rate observed in the HE branch provides insights into exciton dynamics within the LE branch.  
However, at 1.7\,K, acoustic phonon absorption is very unlikely. Therefore, the slower decay is only visible for a few voltages.
Since the optical count rates for this slow decay are very low at 1.7\,K (see Fig.\,\ref{fig:decays_exp}(b)), a precise extraction is not possible.
To validate this physical picture, we conducted lifetime measurements on a different QDM sample at a higher temperature of 10\,K, where acoustic phonon absorption is more pronounced due to the underlying Bose-Einstein statistics.
Fig.\,\ref{app:decay10K} in the Appendix \ref{app:FreddysData} presents this additional measurement, clearly showing the slow decay rate for all measured HE decays.

In the following, we provide a detailed analysis of the mismatch between the radiative decay rates predicted by the effective model and those measured experimentally, attributing the discrepancy to phonon resonances that occur when the $q$-vector of the involved phonon matches a characteristic length scale in the QDM.

\section{Theory}
\label{sec:model}

In this section, we briefly describe the theoretical models utilized in the numerical simulations. The calculation details are presented in the Appendix~\ref{app:calc_details}.
We then continue to discuss the relation between the theoretical and experimental results.

\subsection{The theoretical framework}

We represent the shape of the QDs as truncated Gaussians~\cite{Gawarecki2023}. The model adopted in the calculations is shown in Fig.~\ref{fig:theoretical_model}(a) and inspired by scanning transmission electron microscopy images presented in Ref.~\cite{Lobl2019a}. 

The lattice mismatch between the QDs material (\mbox{InGaAs}) and the GaAs results in strain in the system. We obtain the strain distribution within the continuous elasticity approach~\cite{Pryor1998b} by minimization of the elastic energy. As the lattice constants of AlAs and GaAs are very similar, the AlGaAs barrier is neglected at this stage of simulations. We calculated the piezoelectric potential by taking the strain-induced polarization up to the second order in the strain tensor elements~\cite{Bester2006b}.

The electron and hole single-particle states in the QDM are calculated using the eight-band $\kp$ Hamiltonian~\cite{Winkler2003, Trebin1979} in the envelope function approximation. The details of the implementation and the material parameters are given in the Appendix of Ref.~\onlinecite{Gawarecki2018a}. We account for the Al\(_{0.33}\)Ga\(_{0.67}\)As barrier in the $\kp$ calculation by altering the band gap and valence band offset.  
To visualize the effects of the barrier and strain, we compute the \(\Gamma\)-point band edges by diagonalizing the eight-band Bir-Pikus Hamiltonian for virtual bulk systems with compositions and strain equal to those in the actual system along a line passing through the centers of the QDs (see Fig.~\ref{fig:theoretical_model}(b)).
The single-particle wave functions form a basis for further calculations of the neutral exciton states. The latter has been done within the configuration-interaction (CI) method. At this stage, we also include the electric field (the Stark effect) into the model, as described in the Appendix~\ref{app:calc_details}.

To obtain the radiative relaxation times of the exciton states, we calculate their oscillator strength~\cite{Andrzejewski2010,Gawelczyk2017}. This is performed using the interband momentum matrix elements, determined within the Hellman-Feynman theorem~\cite{Feynman1939, LewYanVoon1993, Eissfeller2012} with the eight-band $\kp$ Hamiltonian.

The phonon-assisted transition rates between the exciton states were calculated within the Fermi golden rule. Here, we take into account the coupling via deformation potential and piezoelectric field~\cite{Krzykowski2020,Woods2004}.
The details are given in appendix.\,\ref{app:calc_details}.
 
\subsection{The system dynamics}

\begin{figure}[tb!!]
    \begin{center}
        \includegraphics[width=.48\textwidth]{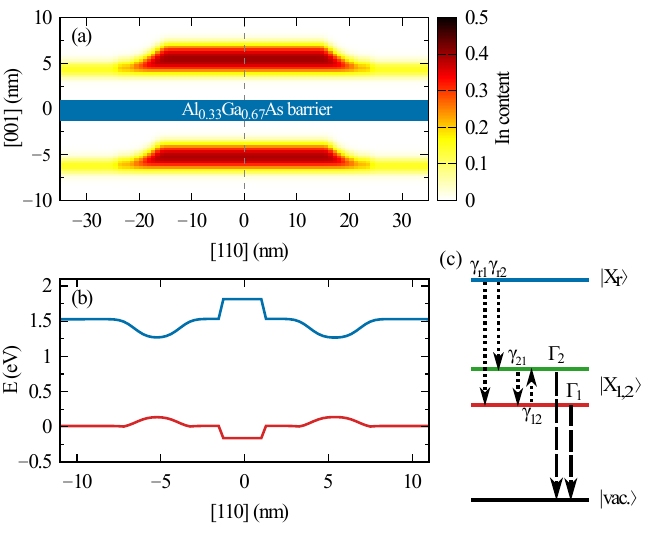}
    \end{center}
    \caption{\label{fig:theoretical_model} (a) The model of In-distribution in the structure and the Al$_{0.33}$Ga$_{0.67}$As barrier. (b) The conduction band and heavy-hole band edges along the axis crossing the QDs centers [as marked in panel (a) by the dashed line].
    (c) Schematic structure of states and processes driving the system dynamics.}
\end{figure}

The calculated quantities constitute a starting point for studying the kinetics of the system.
A schematic structure of the relevant states and processes driving the system is shown in Fig.\,\ref{fig:theoretical_model}(c).
$\ket*{X_\mathrm{1,2}}$ are the HE and LE states, respectively, $\ket*{X_\mathrm{r}}$ is the reservoir of the higher states of the system (the p-shell in the experiment). The states $\ket*{X_\mathrm{1,2}}$ are fed from the reservoir via a very fast decay with the rates $\gamma_{\mathrm{r1}}$ and $\gamma_{\mathrm{r2}}$. This provides an effective description of the states excited via the p-shell, which can be approximately associated with \( p \)-\( p \) excitons. These states decay to the ground (vacuum) state via radiative decay with the rates  $\Gamma_{\mathrm{1}}$ and $\Gamma_{\mathrm{2}}$. In addition, phonon emission and absorption processes induce transitions between these two states with the rates  $\gamma_{\mathrm{12}}$ and $\gamma_{\mathrm{21}}$.
This kinetics is described by the rate equations 
\begin{subequations}
\begin{align}
    \dv{N_0}{t} &= \Gamma_1 N_1 + \Gamma_2 N_2,\label{eq:rate-a} \\
    \dv{N_1}{t} &= \gamma_{31} N_3 + \gamma_{21} N_2 - (\gamma_{12} + \Gamma_1) N_1, \\
    \dv{N_2}{t} &= \gamma_{32} N_3 + \gamma_{12} N_1 - (\gamma_{21} + \Gamma_2) N_2, \\
    \dv{N_\mathrm{r}}{t} &= -(\gamma_{\mathrm{r}1} + \gamma_{\mathrm{r}2}) N_3
    \label{eq:rate-d},
\end{align}	    
\end{subequations}
where $N_0$ is the occupation of the vacuum state; $N_1$, $N_2$ are the occupations of the two lowest exciton states (neglecting the fine structure), and $N_{\mathrm{r}}$ accounts for the occupation of the state representing the reservoir. The set of differential equations is solved numerically, and the resulting occupations are convolved with a Gaussian instrumental response function,
as described in the Appendix~\ref{app:calc_details}.

\subsection{Results and comparison to experiment}
\label{sec:results_theory}

\begin{figure}[tb]
    \begin{center}
        \includegraphics[width=.48\textwidth]{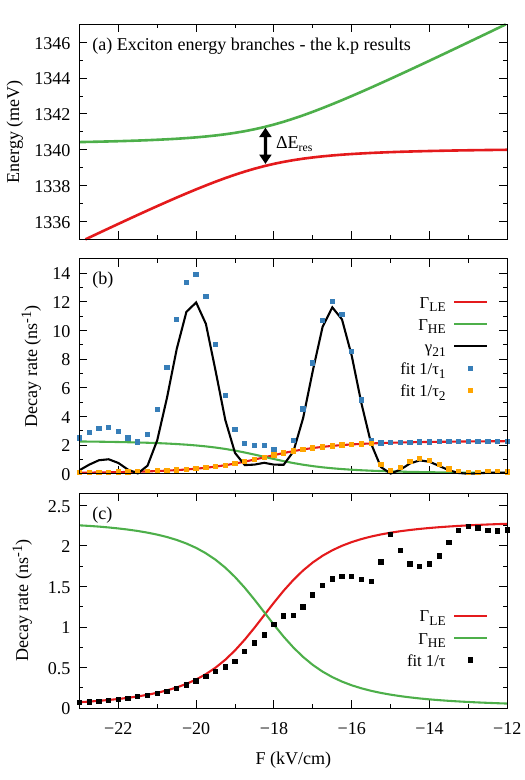}
    \end{center}
    \caption{\label{fig:theory_decay} (a) The theoretical results for the two-lowest exciton energy branches calculated as a function of the axial electric field. The $\Delta E_\mathrm{res}$ denotes the energy difference at the resonance. (b) The electric-field dependence of the relaxation times ($\tau^{-1}_1$ and $\tau^{-1}_2$) fitted to the numerical $\kp$ results via the convoluted biexponential decay. The solid lines denote the radiative recombination rates calculated directly from the exciton states. (c) The decay rates $\tau^{-1}$ were obtained via the convoluted mono-exponential decay.
    }
\end{figure}		

The results of the theoretical model for the exciton energy levels are presented in Fig.~\ref{fig:theory_decay}(a).
We find very good agreement with the experimental data (Fig.~\ref{fig:sample}(b)), with an energy splitting of $\Delta E_\mathrm{res} = 2.01$~meV at the avoided crossing. 

We calculated the radiative recombination rates $\Gamma_n$ and the rates of transitions mediated by the acoustic phonons $\gamma_{nm}$. 
The radiative recombination rates $\Gamma_\mathrm{LE}$ and $\Gamma_\mathrm{HE}$ for the LE and HE branches, obtained from the calculated exciton states (Eq.~\eqref{eq:gamma_th_rad}), are shown in Fig.~\ref{fig:theory_decay}(b) and (c) by the solid red and green lines. As observed from the values far from the avoided crossing, the radiative recombination rate of the direct exciton is $\Gamma_{D} = 2.28$~ns$^{-1}$, which is close to the measured value of $2.45\pm 0.008$~ns$^{-1}$.
We calculated the phonon-assisted relaxation rate from the HE to the LE exciton branch ($\gamma_{21}$) as a function of the electric field. This dependence shows an oscillatory behavior, which is linked to the underlying phonon spectral density~\cite{Wu2014}. The calculated phonon-assisted relaxation rate $\gamma_{21}$ reaches values of up to approximately $12$\,ns$^{-1}$ at $F=-20.0$\,kV/cm and $F=-16.5$\,kV/cm, which is significantly faster than the radiative decay.
These resonances correspond to an energy splitting of the HE and LE branch of $\Delta E \approx 2.75$\,meV.
However, for electric fields far from the avoided crossing, the overlap between the electron wave functions decreases, causing the $\gamma_{21}$ rate to vanish.
The positions of the second maxima in $\gamma_{21}$ are $F=-22.3$\,kV/cm and $F=-14.3$\,kV/cm, both corresponding to an energy splitting of $\Delta E \approx 4.7$\,meV.

To compare the measured decay rates shown in Fig.\,\ref{fig:decays_exp} with our model, we solve the rate equations \eqref{eq:rate-a}--\eqref{eq:rate-d} for the four-level system ($N_{i}(t)$) using the previously calculated values of $\Gamma_{n}$ and $\gamma_{nm}$. The values of $\gamma_{r1}$, $\gamma_{r2}$ were taken from the effective model (see Eq.~\eqref{eq:gamma_reservoir}). We then apply the convolution (Eq.~\eqref{eq:conv}), with the instrument response function $f(t)$ (see Eq. 4), using the same parameter values as in the experimental fitting, obtaining $\widetilde{N}_{i}(t)$.
To ensure a consistent comparison between the exciton dynamics predicted by the full theoretical model and the experimental results, the same fitting scheme must be applied in both cases.

For the HE branch, we fitted $\widetilde{N}_{i}(t)$ using a sum of two exponentials convolved with a Gaussian (see Eq.~\ref{eq:gh} in Appendix\,\ref{app:Decay_fitting}): with the faster ($\tau^{-1}_1$) and with the slower ($\tau^{-1}_2$) decay. The results are shown in Fig.\,\ref{fig:theory_decay}(b). The obtained fast decay rate $\tau^{-1}_1$ follows the behavior of the phonon-assisted relaxation rate $\gamma_{21}$, as expected. The two maxima at $F=-20.0$~kV/cm and $F=-16.5$~kV/cm, corresponding to $\Delta E \approx 2.75$~meV, align very well with the two peaks observed at 0.25\,V and 0.32\,V in the experimental data shown in Fig.\,\ref{fig:decays_exp}(c). This agreement provides clear evidence of experimentally observed phonon resonances. However, in the experiment, the decay rate between the two peaks is not reduced as significantly as predicted by the model. This can be related to the fact that the phonon spectral density crucially depends on the detailed shape of the wave function (hence the specific QD morphology)~\cite{Stock2011a}, and the model approximates the exact material distribution and the geometries of both QDs forming the molecule.

For the LE branch, we fit $\widetilde{N}_{i}(t)$ with a single exponentially modified Gaussian (see Eq.~\ref{eq:gl} in Appendix\,\ref{app:Decay_fitting}). As shown in Fig.~\ref{fig:theory_decay}(c), the obtained decay rate $\tau^{-1}$ is generally consistent with the value of the radiative lifetime $\Gamma_{\mathrm{LE}}$ as expected. However, similarly to the experimental results presented in Fig.~\ref{fig:decays_exp}(d) some deviations appear for the electric fields larger than the value corresponding to the avoided crossing. There are two dips in the theoretical results shown in Fig.~\ref{fig:theory_decay}(c). The region of the plot where these features appear roughly corresponds to the electric field $F$ ranging from $-15.75$~kV/cm to $-13.5$~kV/cm, which corresponds to exciton energy splittings $\Delta E$ from $3.3$~meV to $5.4$~meV. This is consistent with the position of the dip (anti-resonance) observed in the experimental data ($\Delta E =4.8$~meV at 0.36\,V), clearly demonstrating the experimental observation of phonon-related features.
It should be noted that in the theoretical model, phonons are treated assuming the GaAs bulk dispersion.  
However, the actual dispersion in a strained heterostructure may be different, which could lead to a case where only one dip is visible in the experiment.

\section{Conclusions}
\label{sec:conclusions}
In our joint experimental-theoretical work, we present an investigation of phonon-mediated processes in an electrically tunable self-assembled QDM.
Through voltage-dependent measurements, we directly measure the orbital phonon relaxation rates and benchmark our findings against microscopic \kp calculations based on accurate mapping of the QDM morphology.
Our results reveal phonon-mediated relaxation rates that exhibit pronounced resonances and anti-resonances, with rates ranging from several ten ns$^{-1}$ up to tens of $\mu$s$^{-1}$.
Comparison with a kinetic model reveals the voltage (energy) dependent phonon coupling strength and fully explains the interplay between phonon-assisted relaxation and radiative recombination.
The resonances (anti-resonances) observed in the relaxation rates as a function of applied voltage are attributed to enhancement (suppression) of phonon coupling
for specific energies.
These anti-resonances can be leveraged to increase the lifetime of energetically unfavorable charge configurations as required for realizing efficient spin-photon
interfaces and creating multi-dimensional cluster states.
Moreover, suppression of orbital-phonon coupling is likely to be associated with inhibited phonon-mediated decoherence of the quantum states and therefore an essential steps toward future quantum information applications \cite{economouOpticallyGenerated2Dimensional2010, lindnerProposalPulsedOnDemand2009, vezvaee2022}.

\acknowledgments
The authors gratefully acknowledge financial support from
the BMBF for financial support via the QR.N consortium via sub-projects FKZ 16KIS2197 (JJF), 16KIS2200 (AL), 16KIS2193 (SR) and 16KIS2206 (DR) and Germany’s Excellence Strategy (MCQST, EXC-2111, 390814868).
ML and JJF acknowledge funding by the Bavarian Hightech Agenda within the Munich Quantum Valley doctoral fellowship program (ML) and Munich Quantum Valley (JJF).
KG and JJF acknowledge the DAAD-NAWA for financial support via the center-to-center exchange grant 57754510.
KG and PM acknowledge funding from the Polish National Science Centre (NCN) under Grant No. 2016/23/G/ST3/04324. Calculations have been carried out using resources provided by Wroclaw Centre for Networking and Supercomputing (\url{http://wcss.pl}), Grant No.~203.
FB gratefully acknowledges the Exploring Quantum Matter (ExQM) program funded by the State of Bavaria.
AL acknowledges funding by the QuantERA BMBF EQSOTIC project 16KIS2061 as well as the DFG excellence cluster ML4Q project EXC 2004/1.\\

\appendix
\section*{Appendix}

\label{app:appendix}

\subsection{Sample}
\label{app:sample}

The QDM investigated here was fabricated using solid-source molecular beam epitaxy, comprising two vertically stacked indium arsenide (InAs) QDs which are embedded within a gallium arsenide (GaAs) matrix. The heights of the top and bottom QDs were precisely controlled at 2.9\,nm and 2.7\,nm, respectively, through the In-flush technique during the growth process \cite{Wasilewski1990}.
This specific height configuration enables electric tuning of the orbital states in the conduction band when operating the diode in reverse bias \cite{Bracker2006}.
The top dot has a larger height in order to ensure that the hole is always located in this dot.
The separation between the wetting layers of both dots is 10\,nm. An aluminum gallium arsenide (Al$_{0.33}$Ga$_{0.67}$As) barrier with a thickness of 2.5\,nm is placed between the QDs, thereby influencing the coupling strength. Additionally, a 50\,nm thick Al$_{0.33}$Ga$_{0.67}$As tunnel barrier was deposited 5\,nm below the QDM to extend electron tunneling times out of the QDM.

The QDM is integrated into a p-i-n diode structure, with the doped regions serving as electrical contacts for gating purposes.
The diode contacts are situated more than 150\,nm away from the QDM to mitigate uncontrolled charge tunneling into the QDM.
Furthermore, a Al(Ga)As/GaAs distributed Bragg reflector was incorporated below the diode, while a circular Bragg grating was deterministically positioned above a single pre-selected QDM using in-situ electron beam lithography to maximize photon in-coupling and out-coupling efficiencies \cite{Schall2021}.

\subsection{Experimental setup}
\label{app:setup}
All measurements discussed in the main text were performed at a temperature of 1.7\,K within an Attodry2100 dry magnet system, employing a pulsed fs-laser (Coherent Mira 900f) for excitation of the QDM.
A 5\,ps laser pulse is applied to excite the crystal ground state ($\left|\mathrm{vac}\right>$) to the p-orbital of the neutral exciton (p-shell excitation).
The phonon-mediated relaxation process between the p-shell and the s-shell occurs on timescales of a few picoseconds \cite{Narvaez2006, Li1999, Urayama2001}, while the excitonic lifetime ranges from several hundreds of picoseconds to a few nanoseconds \cite{Bacher1999, nakaokaDirectObservationAcoustic2006}. Therefore, the direct radiative emission from the p-shell is weak compared to the s-shell emission we are measuring here.

The occupations of the HE and LE states of the neutral exciton are monitored by detecting the photons emitted by the driven two-level system.
The photons are detected using a time tagger and single photon detectors.
Hereby, we use single-photon avalanche diodes (SPAD) and superconducting nanowire single-photon detectors (SNSPD).
The SPAD has a timing jitter of about 150\,ps.
Since the measured lifetimes are reaching the limit of our SPADs time resolution, the faster decay rates are measured with a SNSPD with a jitter of about 15\,ps.
For the low E decay times below 0.23\,V a pulse picker acousto-optical modulator (AOM) was used to excite the QDM with only every 2nd or 4th laser pulse due to the long lifetime of the indirect exciton states.
All other measurements were performed without a pulse picker AOM.

\subsection{Coupled two-state Hamiltonian}
\label{app:PLVSimulation}

The neutral exciton in our system can be described by the following effective two-state Hamiltonian:

\begin{equation}
    H_{\mathrm{eff}} = \mqty(\varepsilon_\mathrm{I} + \alpha \widetilde{F}& V_\mathrm{t} \\ V_\mathrm{t} & \varepsilon_\mathrm{D}),
    \label{eq:eff}
\end{equation}
where $\varepsilon_\mathrm{I}$, $\varepsilon_\mathrm{D}$ are the energies of the indirect and direct exciton states when the tunnel coupling and external electric field are neglected.
The dependence on the gate-voltage ($\widetilde{F}$) enters via the term $\alpha \widetilde{F}$.
The tunnel coupling is represented by the parameter $V_\mathrm{t}$, which is half of the energy splitting at the resonance $V_\mathrm{t} = \Delta E_\mathrm{res}/2$.
By fitting this to the voltage-dependent photoluminescence spectra for the considered sample in Fig.\,\ref{fig:sample}(b), we obtain the following values: $\varepsilon_\mathrm{I} = 1324.2$~meV, $\varepsilon_\mathrm{D} = 1341.6$~meV, $\alpha = 6.05 \cdot 10^{-2}$~e$\cdot$nm, and $ V_\mathrm{t} =  0.994$~meV.
Our avoided crossing lies at $0.288$~V with an energy splitting of $\Delta E_\mathrm{res} = 1.988$~meV.

\subsection{Fit functions of lifetime measurements}
\label{app:Decay_fitting}

\subsubsection{Exponential convolved with a Gaussian}
The lifetime measurements with a mono-exponential decay are fitted with an exponential function convolved with a Gaussian (exponentially modified Gaussian) supplemented by a constant $C$ accounting for the background
	\begin{equation}
		\label{eq:gl}
		\mathcal{G}_\mathrm{l}(t) = \frac{A \gamma}{2} \, e^{\gamma \qty(\mu - x + {\gamma \sigma^2}/{2} )} \, \mathrm{erfc} \qty(\frac{\mu + \gamma \sigma^2 -x}{\sqrt{2}{\sigma}}) + C,
	\end{equation}
where $\mu$ is the position of the Gaussian component and $\sigma$ its standard deviation. $\gamma = \tau^{-1}$ defines the exponential decay, $A$ is a scaling constant, and erfc() is the complementary error function.
The exponential decay arises from an ideal Lorentzian line-width of the QD emission, while the Gaussian component is caused by an instrument response function~\cite{Zegel1986}.
We obtain $\sigma=0.16$~ns$^{-1}$ for the measurements taken with the SPAD and $\sigma=0.04$~ns$^{-1}$ for the once taken with the SNSPDs.
$\mu$ is the time delay until the first counts are detected and is there different for the measurements done with the SPAD and SNSPD as well as the once with and without pulse picker AOM.

\subsubsection{Sum of two exponentials convolved with a Gaussian}
The lifetime measurements with two exponential components in the decay are fitted by a sum of two exponential functions convolved with a single Gaussian
\begin{align}
    \label{eq:gh}
    \mathcal{G}_\mathrm{h}(t) =& \frac{1}{2} \qty( A \gamma_1 e^{\gamma_1 \qty(\mu - x + {\gamma_1 \sigma^2}/{2} )} + B \gamma_2 e^{\gamma_2 \qty(\mu - x + {\gamma_2 \sigma^2}/{2} )} ) \nonumber \\  & \times \mathrm{erfc} \qty(\frac{\mu + \gamma \sigma^2 -x}{\sqrt{2}{\sigma}}) + C,
\end{align}
where $A$, $B$ represent the scaling factors for the two decays with the $\gamma_1 = \tau_1^{-1}$ and $\gamma_2 = \tau_2^{-1}$ rates, respectively.
All other parameters are defined as above in the mono-exponential case.

\subsection{Decay rates at 10 K}
\label{app:FreddysData}

\begin{figure}[tb]
		\begin{center}
			\includegraphics[width=.48\textwidth]{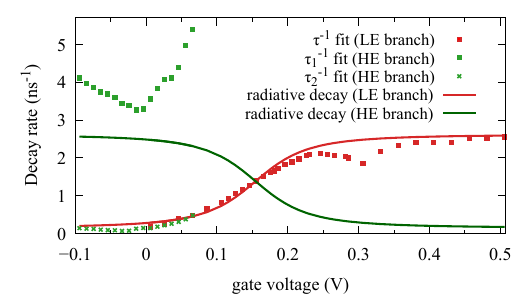}
		\end{center}
		\caption{\label{app:decay10K}  Voltage dependent lifetime measurements at a temperature of 10\,K.
        The fast and slow decay rates of the higher energy branch are shown in green, the decay rates of the lower energy branch in red.
        The solid lines denote the radiative recombination rates calculated from Eq.~\ref{eq:gamma_rad}.
		}
	\end{figure}

The decay rates of the neutral exciton presented in Fig.\,\ref{fig:decays_exp} were measured at 1.7\,K.
A different QDM with the same layer structure was measured at 10\,K exhibiting the same behavior in the voltage dependent decay rates.
Due to the higher temperature, up-scattering is much more likely, leading to a double exponential decay in the high energy emission with a stronger pronounced slow decay.
Fig.\,\ref{app:decay10K} shows the results.

\subsection{Calculation details}
\label{app:calc_details}

\subsubsection{QD model and strain distribution}

The QDs are modeled as truncated Gaussians~\cite{Gawarecki2023}, where their top surfaces are given by the formula
\begin{equation}
    S(x,y) = \mathrm{min}\qty(w \exp{-\frac{\qty( x^2 + y^2)}{l^2}}, h),
\end{equation}
where $w$ is related to the slope steepness, $h$ is the maximal height, and the lateral extension is set by the parameter $l$. For both QDs, we use $l=12$~nm and $w = 15$~nm. The heights are $h_\mathrm{u} = 2.4$~nm and $h_\mathrm{l} = 2.1$~nm (counting from the top of the wetting layer) for the upper and the lower QD, respectively. The thickness of the wetting layer is $0.6$~nm. All the geometric parameters are subjected to the discretization scheme with a mesh size of $a = 0.5653$~nm for the in-plane directions and $a/2$ for the growth direction. The material composition in the QDs and the wetting layers is In$_{0.4}$Ga$_{0.6}$As. To take into account the material intermixing, the InGaAs material distribution is processed by the Gaussian blur with a standard deviation of $0.6$~nm.  Finally, we introduce a $2.3$~nm thick Al$_{0.33}$Ga$_{0.67}$As barrier between the dots.

\subsubsection{Configuration-interaction model}
	
The neutral exciton states are calculated using the configuration-interaction (CI) method within the basis of $24$ lowest electron and 24 lowest hole states. The implementation description is provided in the Supplementary Materials of Ref.~\onlinecite{Ardelt2016}. We take into account the Stark effect due to the axial electric field $F$ via a band-diagonal term
$
V_{\mathrm{F}}(\rr) = - \abs{e} F (z - z_0),
$
where $z_0$ is the center of the computational box. We calculate the matrix elements of $V_{\mathrm{F}}(\rr)$ in the basis of single-particle states and incorporate them into the many-body CI Hamiltonian~\cite{Swiderski2016}.  
\begin{equation}
    \ket{X_n} = \sum_{i,j} c^{(n)}_{ij} a_i^{\dagger} h_j^{\dagger} \ket{\mathrm{vac.}},
\end{equation}
where $\ket{\mathrm{vac.}}$ is the vacuum state, and $a_j^{\dagger}$, $h_j^{\dagger}$ are the creation operators for the electron and hole, respectively. One should note that exciton energy levels form multiplets containing 4 states split via
the exchange interaction~\cite{Bayer2002}. We here consider one bright configuration per multiplet.

\subsubsection{Radiative lifetime}
 
The oscillator strength describing the exciton recombination can be written as~\cite{Gawelczyk2017, Andrzejewski2010}
\begin{equation*}
    f_{n} = -\frac{2}{3 m_0  E^{(\mathrm{X})}_n} \sum_{m=x,y,z}  \abs{ \sum_{i,j} c^{(n)}_{ij} \mel{\Psi^{\mathrm{(v)}}_i}{p_m}{\Psi^{\mathrm{(e)}}_j}}^2,
\end{equation*}
where $E^{(\mathrm{X})}_n$ is the energy corresponding to the $\ket{X_n}$ state, $\Psi^{\mathrm{(v/e)}}_i$ are the single-particle states from the valence/conduction band. The matrix elements of the momentum operator $(p_m)_{ij} \equiv \mel{\Psi^{\mathrm{(v)}}_i}{p_m}{\Psi^{\mathrm{(e)}}_j}$ are calculated using the Hellman-Feynman theorem~\cite{Feynman1939, LewYanVoon1993, Eissfeller2012}.
\begin{equation}
    \label{eq:hf}
    \qty(p_m)_{ij} \approx \frac{m_0}{\hbar} \mel{\Psi^{\mathrm{(v)}}_i}{ \pdv{H_{\bm{k}\cdot\bm{p} }}{k_m} }{\Psi^{\mathrm{(e)}}_j},
\end{equation}
where $H_{\bm{k}\cdot\bm{p}}$ is the eight-band $\kp$ Hamiltonian. The radiative recombination rate for the state $\ket{X_n}$ can be calculated from~\cite{Thranhardt2002,Gawelczyk2017}
\begin{equation}
    \label{eq:gamma_th_rad}
    \Gamma_n =  f_{n0} \qty(E^{(\mathrm{X})}_n)^2 \frac{n_\mathrm{r} e^2 }{ 6 \pi \epsilon_0 \hbar^2 m_0 c^3},
\end{equation}
where $n_\mathrm{r}$ is the refractive index, $\epsilon_0$ is the vacuum permittivity, and $c$ is the speed of light. For GaAs we use $n_\mathrm{r} = 3.347$ ~\cite{Amirtharaj1994}.

\subsubsection{The exciton-phonon interaction}
The interactions between the carriers and acoustic phonons are described by the Hamiltonian $H_\mathrm{int}$, which can be expressed by terms involving individual phonon modes~\cite{Krzykowski2020}
\begin{align*}	
    H_\mathrm{int} = \sum_{\bm{q}, \lambda} \mathcal{V}(\bm{q},\lambda) e^{i \bm{q} \bm{r}},
\end{align*}
where $\lambda$ are phonon branches (one longitudinal and two transversal) and $\bm{q}$ are the phonon wave vectors. In $\mathcal{V}(\bm{q},\lambda)$, we take into account the contributions stemming from the deformation-potential coupling as well as the piezoelectric-field coupling. The complete description of the model can be found, e.g., in Ref.~\onlinecite{Krzykowski2020}. To calculate the phonon-assisted transitions between the exciton states $\ket{X_n} \rightarrow \ket{X_m}$, we utilize the Fermi golden rule
\begin{align}
    \label{eq:gamma_th_ph}
    \gamma_{nm} =& \frac{2 \pi}{ \hbar^2} \abs{n_B({\omega_{mn}}) + 1} \sum_{\bm{q}, \lambda} \abs{F^{\mathrm{(X)}}_{nm,\lambda}(\bm{q})}^2 \nonumber \\ &  \times \qty[ \delta\qty({\omega_{mn} - c_\lambda q}) + \delta\qty({\omega_{mn} + c_\lambda q}) ],
\end{align}
where $n_B(\omega)$ is the Bose-Einstein distribution, $\omega_{ij} = \Delta E^{\mathrm{(X)}}_{ij}/\hbar = \qty(E^{\mathrm{(X)}}_j - E^{\mathrm{(X)}}_i)/\hbar$, and $c_{\lambda}$ is the phonon-branch-dependent speed of sound. The exciton form-factor $F^{\mathrm{(X)}}_{nm,\lambda}(\bm{q})$ is defined in terms of the electron $F^{\mathrm{(e)}}_{nm,\lambda}(\bm{q})$ and hole $F^{\mathrm{(h)}}_{nm,\lambda}(\bm{q})$ form-factors~\cite{Daniels2013}
\begin{align*}
    F^{\mathrm{(X)}}_{nm,\lambda}(\bm{q}) = \sum_{i,j,i',j'} c^{(n)*}_{ij} c^{(m)}_{i'j'} \qty( F^{\mathrm{(e)}}_{ii',\lambda}(\bm{q}) \delta_{jj'} + F^{\mathrm{(h)}}_{jj',\lambda}(\bm{q}) \delta_{ii'}),
\end{align*}
where
\begin{align*}
    F^{\mathrm{(e/h)}}_{ij,\lambda}(\bm{q}) = \mel{\Psi^{\mathrm{(e/h)}}_i}{\mathcal{V}(\bm{q},\lambda) e^{i \bm{q} \bm{r}}}{\Psi^{\mathrm{(e/h)}}_j}.
\end{align*}

\subsubsection{The rate equations}

The relaxation rates from the reservoir $\gamma_{\mathrm{r},1}$, $\gamma_{\mathrm{r},2}$ that enter the rate equations  (${\dd N_i}/{ \dd t}$) are calculated from
\begin{equation}
    \label{eq:gamma_reservoir}
    \gamma_{\mathrm{r} n}(\widetilde{F}) = \abs{c^{(\mathrm{D})}_n (\widetilde{F})}^2 \gamma^{(\mathrm{D})}_\mathrm{r} + \abs{c^{(\mathrm{I})}_n (\widetilde{F})}^2 \gamma^{(\mathrm{I})}_\mathrm{r},
\end{equation}
where $c^{(\mathrm{D})}_n (\widetilde{F})$, $c^{(\mathrm{I})}_n (\widetilde{F})$ are the effecive-model coefficients as in Eq.~\eqref{eq:gamma_rad}, and $\gamma^{(\mathrm{D})}_\mathrm{r}$ ($\gamma^{(\mathrm{I})}_\mathrm{r}$) is the intradot (interdot) transition rate between the $p$- and the $s$-type states. We took $\gamma^{(\mathrm{D})}_\mathrm{r} = 160$~ns$^{-1}$ and $\gamma^{(\mathrm{I})}_\mathrm{r} = 20$~ns$^{-1}$ owing to the fact that intradot relaxation is much faster than phonon-assisted tunneling.

We numerically solve the set of differential equations using Scipy~\cite{2020SciPy-NMeth} library (\texttt{odeint}) to obtain the time evolution of the occupations $N_i(t)$.  For comparison with the experimental data, the results are convolved with the instrument response function $f(t)$~\cite{Zegel1986}
\begin{equation}
    \label{eq:conv}
    \widetilde{N}_i(t) = \int_{-\infty}^{\infty} f(\tau) \Theta(t-\tau)  \, N_i(t-\tau) \dd{\tau},
\end{equation}
where $\Theta(t)$ is the Heaviside step function and
\begin{equation*}
    f(t) = \frac{1}{\sigma \sqrt{2 \pi}} \exp(-\frac{(t-\mu)^2}{2 \sigma^2}).
\end{equation*}
is the Gaussian centered at $\mu$.
We use the same parameter values as those in Sec.~\ref{sec:exp}.

\bibliography{main}

\begin{thebibliography}{60}%
\makeatletter
\providecommand \@ifxundefined [1]{%
 \@ifx{#1\undefined}
}%
\providecommand \@ifnum [1]{%
 \ifnum #1\expandafter \@firstoftwo
 \else \expandafter \@secondoftwo
 \fi
}%
\providecommand \@ifx [1]{%
 \ifx #1\expandafter \@firstoftwo
 \else \expandafter \@secondoftwo
 \fi
}%
\providecommand \natexlab [1]{#1}%
\providecommand \enquote  [1]{``#1''}%
\providecommand \bibnamefont  [1]{#1}%
\providecommand \bibfnamefont [1]{#1}%
\providecommand \citenamefont [1]{#1}%
\providecommand \href@noop [0]{\@secondoftwo}%
\providecommand \href [0]{\begingroup \@sanitize@url \@href}%
\providecommand \@href[1]{\@@startlink{#1}\@@href}%
\providecommand \@@href[1]{\endgroup#1\@@endlink}%
\providecommand \@sanitize@url [0]{\catcode `\\12\catcode `\$12\catcode
  `\&12\catcode `\#12\catcode `\^12\catcode `\_12\catcode `\%12\relax}%
\providecommand \@@startlink[1]{}%
\providecommand \@@endlink[0]{}%
\providecommand \url  [0]{\begingroup\@sanitize@url \@url }%
\providecommand \@url [1]{\endgroup\@href {#1}{\urlprefix }}%
\providecommand \urlprefix  [0]{URL }%
\providecommand \Eprint [0]{\href }%
\providecommand \doibase [0]{https://doi.org/}%
\providecommand \selectlanguage [0]{\@gobble}%
\providecommand \bibinfo  [0]{\@secondoftwo}%
\providecommand \bibfield  [0]{\@secondoftwo}%
\providecommand \translation [1]{[#1]}%
\providecommand \BibitemOpen [0]{}%
\providecommand \bibitemStop [0]{}%
\providecommand \bibitemNoStop [0]{.\EOS\space}%
\providecommand \EOS [0]{\spacefactor3000\relax}%
\providecommand \BibitemShut  [1]{\csname bibitem#1\endcsname}%
\let\auto@bib@innerbib\@empty
\bibitem [{\citenamefont {Awschalom}\ \emph {et~al.}(2018)\citenamefont
  {Awschalom}, \citenamefont {Hanson}, \citenamefont {Wrachtrup},\ and\
  \citenamefont {Zhou}}]{Awschalom2018}%
  \BibitemOpen
  \bibfield  {author} {\bibinfo {author} {\bibfnamefont {D.~D.}\ \bibnamefont
  {Awschalom}}, \bibinfo {author} {\bibfnamefont {R.}~\bibnamefont {Hanson}},
  \bibinfo {author} {\bibfnamefont {J.}~\bibnamefont {Wrachtrup}},\ and\
  \bibinfo {author} {\bibfnamefont {B.~B.}\ \bibnamefont {Zhou}},\ }\bibfield
  {title} {\bibinfo {title} {Quantum technologies with optically interfaced
  solid-state spins},\ }\href {https://doi.org/10.1038/s41566-018-0232-2}
  {\bibfield  {journal} {\bibinfo  {journal} {Nat. Photon.}\ }\textbf {\bibinfo
  {volume} {12}},\ \bibinfo {pages} {516–527} (\bibinfo {year}
  {2018})}\BibitemShut {NoStop}%
\bibitem [{\citenamefont {Bayer}\ \emph {et~al.}(2002)\citenamefont {Bayer},
  \citenamefont {Ortner}, \citenamefont {Stern}, \citenamefont {Kuther},
  \citenamefont {Gorbunov}, \citenamefont {Forchel}, \citenamefont {Hawrylak},
  \citenamefont {Fafard}, \citenamefont {Hinzer}, \citenamefont {Reinecke},
  \citenamefont {Walck}, \citenamefont {Reithmaier}, \citenamefont {Klopf},\
  and\ \citenamefont {Sch\"afer}}]{Bayer2002}%
  \BibitemOpen
  \bibfield  {author} {\bibinfo {author} {\bibfnamefont {M.}~\bibnamefont
  {Bayer}}, \bibinfo {author} {\bibfnamefont {G.}~\bibnamefont {Ortner}},
  \bibinfo {author} {\bibfnamefont {O.}~\bibnamefont {Stern}}, \bibinfo
  {author} {\bibfnamefont {A.}~\bibnamefont {Kuther}}, \bibinfo {author}
  {\bibfnamefont {A.~A.}\ \bibnamefont {Gorbunov}}, \bibinfo {author}
  {\bibfnamefont {A.}~\bibnamefont {Forchel}}, \bibinfo {author} {\bibfnamefont
  {P.}~\bibnamefont {Hawrylak}}, \bibinfo {author} {\bibfnamefont
  {S.}~\bibnamefont {Fafard}}, \bibinfo {author} {\bibfnamefont
  {K.}~\bibnamefont {Hinzer}}, \bibinfo {author} {\bibfnamefont {T.~L.}\
  \bibnamefont {Reinecke}}, \bibinfo {author} {\bibfnamefont {S.~N.}\
  \bibnamefont {Walck}}, \bibinfo {author} {\bibfnamefont {J.~P.}\ \bibnamefont
  {Reithmaier}}, \bibinfo {author} {\bibfnamefont {F.}~\bibnamefont {Klopf}},\
  and\ \bibinfo {author} {\bibfnamefont {F.}~\bibnamefont {Sch\"afer}},\
  }\bibfield  {title} {\bibinfo {title} {{Fine structure of neutral and charged
  excitons in self-assembled In(Ga)As/(Al)GaAs quantum dots}},\ }\href
  {https://doi.org/10.1103/PhysRevB.65.195315} {\bibfield  {journal} {\bibinfo
  {journal} {Phys. Rev. B}\ }\textbf {\bibinfo {volume} {65}},\ \bibinfo
  {pages} {195315} (\bibinfo {year} {2002})}\BibitemShut {NoStop}%
\bibitem [{\citenamefont {Gao}\ \emph {et~al.}(2012)\citenamefont {Gao},
  \citenamefont {Fallahi}, \citenamefont {Togan}, \citenamefont
  {Miguel-Sanchez},\ and\ \citenamefont {Imamoglu}}]{Gao2012}%
  \BibitemOpen
  \bibfield  {author} {\bibinfo {author} {\bibfnamefont {W.~B.}\ \bibnamefont
  {Gao}}, \bibinfo {author} {\bibfnamefont {P.}~\bibnamefont {Fallahi}},
  \bibinfo {author} {\bibfnamefont {E.}~\bibnamefont {Togan}}, \bibinfo
  {author} {\bibfnamefont {J.}~\bibnamefont {Miguel-Sanchez}},\ and\ \bibinfo
  {author} {\bibfnamefont {A.}~\bibnamefont {Imamoglu}},\ }\bibfield  {title}
  {\bibinfo {title} {Observation of entanglement between a quantum dot spin and
  a single photon},\ }\href {https://doi.org/10.1038/nature11573} {\bibfield
  {journal} {\bibinfo  {journal} {Nature}\ }\textbf {\bibinfo {volume} {491}},\
  \bibinfo {pages} {426–430} (\bibinfo {year} {2012})}\BibitemShut {NoStop}%
\bibitem [{\citenamefont {Stockill}\ \emph {et~al.}(2017)\citenamefont
  {Stockill}, \citenamefont {Stanley}, \citenamefont {Huthmacher},
  \citenamefont {Clarke}, \citenamefont {Hugues}, \citenamefont {Miller},
  \citenamefont {Matthiesen}, \citenamefont {Le~Gall},\ and\ \citenamefont
  {Atat\"ure}}]{Stockill2017}%
  \BibitemOpen
  \bibfield  {author} {\bibinfo {author} {\bibfnamefont {R.}~\bibnamefont
  {Stockill}}, \bibinfo {author} {\bibfnamefont {M.~J.}\ \bibnamefont
  {Stanley}}, \bibinfo {author} {\bibfnamefont {L.}~\bibnamefont {Huthmacher}},
  \bibinfo {author} {\bibfnamefont {E.}~\bibnamefont {Clarke}}, \bibinfo
  {author} {\bibfnamefont {M.}~\bibnamefont {Hugues}}, \bibinfo {author}
  {\bibfnamefont {A.~J.}\ \bibnamefont {Miller}}, \bibinfo {author}
  {\bibfnamefont {C.}~\bibnamefont {Matthiesen}}, \bibinfo {author}
  {\bibfnamefont {C.}~\bibnamefont {Le~Gall}},\ and\ \bibinfo {author}
  {\bibfnamefont {M.}~\bibnamefont {Atat\"ure}},\ }\bibfield  {title} {\bibinfo
  {title} {Phase-tuned entangled state generation between distant spin
  qubits},\ }\href {https://doi.org/10.1103/PhysRevLett.119.010503} {\bibfield
  {journal} {\bibinfo  {journal} {Phys. Rev. Lett.}\ }\textbf {\bibinfo
  {volume} {119}},\ \bibinfo {pages} {010503} (\bibinfo {year}
  {2017})}\BibitemShut {NoStop}%
\bibitem [{\citenamefont {Favero}\ \emph {et~al.}(2003)\citenamefont {Favero},
  \citenamefont {Cassabois}, \citenamefont {Ferreira}, \citenamefont {Darson},
  \citenamefont {Voisin}, \citenamefont {Tignon}, \citenamefont {Delalande},
  \citenamefont {Bastard}, \citenamefont {Roussignol},\ and\ \citenamefont
  {Gerard}}]{Favero2003AcousticPS}%
  \BibitemOpen
  \bibfield  {author} {\bibinfo {author} {\bibfnamefont {I.}~\bibnamefont
  {Favero}}, \bibinfo {author} {\bibfnamefont {G.}~\bibnamefont {Cassabois}},
  \bibinfo {author} {\bibfnamefont {R.}~\bibnamefont {Ferreira}}, \bibinfo
  {author} {\bibfnamefont {D.}~\bibnamefont {Darson}}, \bibinfo {author}
  {\bibfnamefont {C.}~\bibnamefont {Voisin}}, \bibinfo {author} {\bibfnamefont
  {J.}~\bibnamefont {Tignon}}, \bibinfo {author} {\bibfnamefont
  {C.}~\bibnamefont {Delalande}}, \bibinfo {author} {\bibfnamefont
  {G.}~\bibnamefont {Bastard}}, \bibinfo {author} {\bibfnamefont
  {P.}~\bibnamefont {Roussignol}},\ and\ \bibinfo {author} {\bibfnamefont
  {J.~M.}\ \bibnamefont {Gerard}},\ }\bibfield  {title} {\bibinfo {title}
  {{Acoustic phonon sidebands in the emission line of single InAs/GaAs quantum
  dots}},\ }\href {https://api.semanticscholar.org/CorpusID:120242728}
  {\bibfield  {journal} {\bibinfo  {journal} {Physical Review B}\ }\textbf
  {\bibinfo {volume} {68}},\ \bibinfo {pages} {233301} (\bibinfo {year}
  {2003})}\BibitemShut {NoStop}%
\bibitem [{\citenamefont {Kuhlmann}\ \emph {et~al.}(2015)\citenamefont
  {Kuhlmann}, \citenamefont {Prechtel}, \citenamefont {Houel}, \citenamefont
  {Ludwig}, \citenamefont {Reuter}, \citenamefont {Wieck},\ and\ \citenamefont
  {Warburton}}]{Kuhlmann2015}%
  \BibitemOpen
  \bibfield  {author} {\bibinfo {author} {\bibfnamefont {A.~V.}\ \bibnamefont
  {Kuhlmann}}, \bibinfo {author} {\bibfnamefont {J.~H.}\ \bibnamefont
  {Prechtel}}, \bibinfo {author} {\bibfnamefont {J.}~\bibnamefont {Houel}},
  \bibinfo {author} {\bibfnamefont {A.}~\bibnamefont {Ludwig}}, \bibinfo
  {author} {\bibfnamefont {D.}~\bibnamefont {Reuter}}, \bibinfo {author}
  {\bibfnamefont {A.~D.}\ \bibnamefont {Wieck}},\ and\ \bibinfo {author}
  {\bibfnamefont {R.~J.}\ \bibnamefont {Warburton}},\ }\bibfield  {title}
  {\bibinfo {title} {Transform-limited single photons from a single quantum
  dot},\ }\href {https://doi.org/10.1038/ncomms9204} {\bibfield  {journal}
  {\bibinfo  {journal} {Nat. Commun.}\ }\textbf {\bibinfo {volume} {6}},\
  \bibinfo {pages} {8204} (\bibinfo {year} {2015})}\BibitemShut {NoStop}%
\bibitem [{\citenamefont {Lindner}\ and\ \citenamefont
  {Rudolph}(2009)}]{lindnerProposalPulsedOnDemand2009}%
  \BibitemOpen
  \bibfield  {author} {\bibinfo {author} {\bibfnamefont {N.~H.}\ \bibnamefont
  {Lindner}}\ and\ \bibinfo {author} {\bibfnamefont {T.}~\bibnamefont
  {Rudolph}},\ }\bibfield  {title} {\bibinfo {title} {Proposal for {{Pulsed
  On-Demand Sources}} of {{Photonic Cluster State Strings}}},\ }\href
  {https://doi.org/10.1103/PhysRevLett.103.113602} {\bibfield  {journal}
  {\bibinfo  {journal} {Phys. Rev. Lett.}\ }\textbf {\bibinfo {volume} {103}},\
  \bibinfo {pages} {113602} (\bibinfo {year} {2009})}\BibitemShut {NoStop}%
\bibitem [{\citenamefont {Cogan}\ \emph {et~al.}(2023)\citenamefont {Cogan},
  \citenamefont {Su}, \citenamefont {Kenneth},\ and\ \citenamefont
  {Gershoni}}]{Cogan2023}%
  \BibitemOpen
  \bibfield  {author} {\bibinfo {author} {\bibfnamefont {D.}~\bibnamefont
  {Cogan}}, \bibinfo {author} {\bibfnamefont {Z.-E.}\ \bibnamefont {Su}},
  \bibinfo {author} {\bibfnamefont {O.}~\bibnamefont {Kenneth}},\ and\ \bibinfo
  {author} {\bibfnamefont {D.}~\bibnamefont {Gershoni}},\ }\bibfield  {title}
  {\bibinfo {title} {Deterministic generation of indistinguishable photons in a
  cluster state},\ }\href
  {https://doi.org/https://doi.org/10.1038/s41566-022-01152-2} {\bibfield
  {journal} {\bibinfo  {journal} {Nature Photonics}\ }\textbf {\bibinfo
  {volume} {17}},\ \bibinfo {pages} {324} (\bibinfo {year} {2023})}\BibitemShut
  {NoStop}%
\bibitem [{\citenamefont {Tran}\ \emph {et~al.}(2022)\citenamefont {Tran},
  \citenamefont {Bracker}, \citenamefont {Yakes}, \citenamefont {Grim},\ and\
  \citenamefont {Carter}}]{tranEnhancedSpinCoherence2022}%
  \BibitemOpen
  \bibfield  {author} {\bibinfo {author} {\bibfnamefont {K.~X.}\ \bibnamefont
  {Tran}}, \bibinfo {author} {\bibfnamefont {A.~S.}\ \bibnamefont {Bracker}},
  \bibinfo {author} {\bibfnamefont {M.~K.}\ \bibnamefont {Yakes}}, \bibinfo
  {author} {\bibfnamefont {J.~Q.}\ \bibnamefont {Grim}},\ and\ \bibinfo
  {author} {\bibfnamefont {S.~G.}\ \bibnamefont {Carter}},\ }\bibfield  {title}
  {\bibinfo {title} {Enhanced {{Spin Coherence}} of a {{Self-Assembled Quantum
  Dot Molecule}} at the {{Optimal Electrical Bias}}},\ }\href
  {https://doi.org/10.1103/PhysRevLett.129.027403} {\bibfield  {journal}
  {\bibinfo  {journal} {Phys. Rev. Lett.}\ }\textbf {\bibinfo {volume} {129}},\
  \bibinfo {pages} {027403} (\bibinfo {year} {2022})}\BibitemShut {NoStop}%
\bibitem [{\citenamefont {Doty}\ \emph {et~al.}(2010)\citenamefont {Doty},
  \citenamefont {Climente}, \citenamefont {Greilich}, \citenamefont {Yakes},
  \citenamefont {Bracker},\ and\ \citenamefont {Gammon}}]{Doty2010}%
  \BibitemOpen
  \bibfield  {author} {\bibinfo {author} {\bibfnamefont {M.~F.}\ \bibnamefont
  {Doty}}, \bibinfo {author} {\bibfnamefont {J.~I.}\ \bibnamefont {Climente}},
  \bibinfo {author} {\bibfnamefont {A.}~\bibnamefont {Greilich}}, \bibinfo
  {author} {\bibfnamefont {M.}~\bibnamefont {Yakes}}, \bibinfo {author}
  {\bibfnamefont {A.~S.}\ \bibnamefont {Bracker}},\ and\ \bibinfo {author}
  {\bibfnamefont {D.}~\bibnamefont {Gammon}},\ }\bibfield  {title} {\bibinfo
  {title} {Hole-spin mixing in inas quantum dot molecules},\ }\href
  {https://doi.org/10.1103/PhysRevB.81.035308} {\bibfield  {journal} {\bibinfo
  {journal} {Phys. Rev. B}\ }\textbf {\bibinfo {volume} {81}},\ \bibinfo
  {pages} {035308} (\bibinfo {year} {2010})}\BibitemShut {NoStop}%
\bibitem [{\citenamefont {Ardelt}\ \emph {et~al.}(2016)\citenamefont {Ardelt},
  \citenamefont {Gawarecki}, \citenamefont {M\"uller}, \citenamefont {Waeber},
  \citenamefont {Bechtold}, \citenamefont {Oberhofer}, \citenamefont {Daniels},
  \citenamefont {Klotz}, \citenamefont {Bichler}, \citenamefont {Kuhn},
  \citenamefont {Krenner}, \citenamefont {Machnikowski},\ and\ \citenamefont
  {Finley}}]{Ardelt2016}%
  \BibitemOpen
  \bibfield  {author} {\bibinfo {author} {\bibfnamefont {P.-L.}\ \bibnamefont
  {Ardelt}}, \bibinfo {author} {\bibfnamefont {K.}~\bibnamefont {Gawarecki}},
  \bibinfo {author} {\bibfnamefont {K.}~\bibnamefont {M\"uller}}, \bibinfo
  {author} {\bibfnamefont {A.~M.}\ \bibnamefont {Waeber}}, \bibinfo {author}
  {\bibfnamefont {A.}~\bibnamefont {Bechtold}}, \bibinfo {author}
  {\bibfnamefont {K.}~\bibnamefont {Oberhofer}}, \bibinfo {author}
  {\bibfnamefont {J.~M.}\ \bibnamefont {Daniels}}, \bibinfo {author}
  {\bibfnamefont {F.}~\bibnamefont {Klotz}}, \bibinfo {author} {\bibfnamefont
  {M.}~\bibnamefont {Bichler}}, \bibinfo {author} {\bibfnamefont
  {T.}~\bibnamefont {Kuhn}}, \bibinfo {author} {\bibfnamefont {H.~J.}\
  \bibnamefont {Krenner}}, \bibinfo {author} {\bibfnamefont {P.}~\bibnamefont
  {Machnikowski}},\ and\ \bibinfo {author} {\bibfnamefont {J.~J.}\ \bibnamefont
  {Finley}},\ }\bibfield  {title} {\bibinfo {title} {Coulomb mediated
  hybridization of excitons in coupled quantum dots},\ }\href
  {https://doi.org/10.1103/PhysRevLett.116.077401} {\bibfield  {journal}
  {\bibinfo  {journal} {Phys. Rev. Lett.}\ }\textbf {\bibinfo {volume} {116}},\
  \bibinfo {pages} {077401} (\bibinfo {year} {2016})}\BibitemShut {NoStop}%
\bibitem [{\citenamefont {Bopp}\ \emph
  {et~al.}(2023{\natexlab{a}})\citenamefont {Bopp}, \citenamefont {Schall},
  \citenamefont {Bart}, \citenamefont {V\"ogl}, \citenamefont {Cullip},
  \citenamefont {Sbresny}, \citenamefont {Boos}, \citenamefont {Thalacker},
  \citenamefont {Lienhart}, \citenamefont {Rodt}, \citenamefont {Reuter},
  \citenamefont {Ludwig}, \citenamefont {Wieck}, \citenamefont {Reitzenstein},
  \citenamefont {M\"uller},\ and\ \citenamefont {Finley}}]{Bopp2023a}%
  \BibitemOpen
  \bibfield  {author} {\bibinfo {author} {\bibfnamefont {F.}~\bibnamefont
  {Bopp}}, \bibinfo {author} {\bibfnamefont {J.}~\bibnamefont {Schall}},
  \bibinfo {author} {\bibfnamefont {N.}~\bibnamefont {Bart}}, \bibinfo {author}
  {\bibfnamefont {F.}~\bibnamefont {V\"ogl}}, \bibinfo {author} {\bibfnamefont
  {C.}~\bibnamefont {Cullip}}, \bibinfo {author} {\bibfnamefont
  {F.}~\bibnamefont {Sbresny}}, \bibinfo {author} {\bibfnamefont
  {K.}~\bibnamefont {Boos}}, \bibinfo {author} {\bibfnamefont {C.}~\bibnamefont
  {Thalacker}}, \bibinfo {author} {\bibfnamefont {M.}~\bibnamefont {Lienhart}},
  \bibinfo {author} {\bibfnamefont {S.}~\bibnamefont {Rodt}}, \bibinfo {author}
  {\bibfnamefont {D.}~\bibnamefont {Reuter}}, \bibinfo {author} {\bibfnamefont
  {A.}~\bibnamefont {Ludwig}}, \bibinfo {author} {\bibfnamefont {A.~D.}\
  \bibnamefont {Wieck}}, \bibinfo {author} {\bibfnamefont {S.}~\bibnamefont
  {Reitzenstein}}, \bibinfo {author} {\bibfnamefont {K.}~\bibnamefont
  {M\"uller}},\ and\ \bibinfo {author} {\bibfnamefont {J.~J.}\ \bibnamefont
  {Finley}},\ }\bibfield  {title} {\bibinfo {title} {Coherent driving of direct
  and indirect excitons in a quantum dot molecule},\ }\href
  {https://doi.org/10.1103/PhysRevB.107.165426} {\bibfield  {journal} {\bibinfo
   {journal} {Phys. Rev. B}\ }\textbf {\bibinfo {volume} {107}},\ \bibinfo
  {pages} {165426} (\bibinfo {year} {2023}{\natexlab{a}})}\BibitemShut
  {NoStop}%
\bibitem [{\citenamefont {Bopp}\ \emph {et~al.}(2022)\citenamefont {Bopp},
  \citenamefont {Rojas}, \citenamefont {Revenga}, \citenamefont {Riedl},
  \citenamefont {Sbresny}, \citenamefont {Boos}, \citenamefont {Simmet},
  \citenamefont {Ahmadi}, \citenamefont {Gershoni}, \citenamefont {Kasprzak},
  \citenamefont {Ludwig}, \citenamefont {Reitzenstein}, \citenamefont {Wieck},
  \citenamefont {Reuter}, \citenamefont {M{\"u}ller},\ and\ \citenamefont
  {Finley}}]{boppQuantumDotMolecule2022}%
  \BibitemOpen
  \bibfield  {author} {\bibinfo {author} {\bibfnamefont {F.}~\bibnamefont
  {Bopp}}, \bibinfo {author} {\bibfnamefont {J.}~\bibnamefont {Rojas}},
  \bibinfo {author} {\bibfnamefont {N.}~\bibnamefont {Revenga}}, \bibinfo
  {author} {\bibfnamefont {H.}~\bibnamefont {Riedl}}, \bibinfo {author}
  {\bibfnamefont {F.}~\bibnamefont {Sbresny}}, \bibinfo {author} {\bibfnamefont
  {K.}~\bibnamefont {Boos}}, \bibinfo {author} {\bibfnamefont {T.}~\bibnamefont
  {Simmet}}, \bibinfo {author} {\bibfnamefont {A.}~\bibnamefont {Ahmadi}},
  \bibinfo {author} {\bibfnamefont {D.}~\bibnamefont {Gershoni}}, \bibinfo
  {author} {\bibfnamefont {J.}~\bibnamefont {Kasprzak}}, \bibinfo {author}
  {\bibfnamefont {A.}~\bibnamefont {Ludwig}}, \bibinfo {author} {\bibfnamefont
  {S.}~\bibnamefont {Reitzenstein}}, \bibinfo {author} {\bibfnamefont
  {A.}~\bibnamefont {Wieck}}, \bibinfo {author} {\bibfnamefont
  {D.}~\bibnamefont {Reuter}}, \bibinfo {author} {\bibfnamefont
  {K.}~\bibnamefont {M{\"u}ller}},\ and\ \bibinfo {author} {\bibfnamefont
  {J.~J.}\ \bibnamefont {Finley}},\ }\bibfield  {title} {\bibinfo {title}
  {Quantum {{Dot Molecule Devices}} with {{Optical Control}} of {{Charge
  Status}} and {{Electronic Control}} of {{Coupling}}},\ }\href
  {https://doi.org/10.1002/qute.202200049} {\bibfield  {journal} {\bibinfo
  {journal} {Adv Quantum Tech}\ }\textbf {\bibinfo {volume} {5}},\ \bibinfo
  {pages} {2200049} (\bibinfo {year} {2022})}\BibitemShut {NoStop}%
\bibitem [{\citenamefont {Economou}\ \emph {et~al.}(2010)\citenamefont
  {Economou}, \citenamefont {Lindner},\ and\ \citenamefont
  {Rudolph}}]{economouOpticallyGenerated2Dimensional2010}%
  \BibitemOpen
  \bibfield  {author} {\bibinfo {author} {\bibfnamefont {S.~E.}\ \bibnamefont
  {Economou}}, \bibinfo {author} {\bibfnamefont {N.}~\bibnamefont {Lindner}},\
  and\ \bibinfo {author} {\bibfnamefont {T.}~\bibnamefont {Rudolph}},\
  }\bibfield  {title} {\bibinfo {title} {Optically {{Generated}}
  2-{{Dimensional Photonic Cluster State}} from {{Coupled Quantum Dots}}},\
  }\href {https://doi.org/10.1103/PhysRevLett.105.093601} {\bibfield  {journal}
  {\bibinfo  {journal} {Phys. Rev. Lett.}\ }\textbf {\bibinfo {volume} {105}},\
  \bibinfo {pages} {093601} (\bibinfo {year} {2010})}\BibitemShut {NoStop}%
\bibitem [{\citenamefont {Vezvaee}\ \emph {et~al.}(2022)\citenamefont
  {Vezvaee}, \citenamefont {Hilaire}, \citenamefont {Doty},\ and\ \citenamefont
  {Economou}}]{vezvaee2022}%
  \BibitemOpen
  \bibfield  {author} {\bibinfo {author} {\bibfnamefont {A.}~\bibnamefont
  {Vezvaee}}, \bibinfo {author} {\bibfnamefont {P.}~\bibnamefont {Hilaire}},
  \bibinfo {author} {\bibfnamefont {M.~F.}\ \bibnamefont {Doty}},\ and\
  \bibinfo {author} {\bibfnamefont {S.~E.}\ \bibnamefont {Economou}},\
  }\bibfield  {title} {\bibinfo {title} {Deterministic {{Generation}} of
  {{Entangled Photonic Cluster States}} from {{Quantum Dot Molecules}}},\
  }\href {https://doi.org/10.1103/PhysRevApplied.18.L061003} {\bibfield
  {journal} {\bibinfo  {journal} {Phys. Rev. Applied}\ }\textbf {\bibinfo
  {volume} {18}},\ \bibinfo {pages} {L061003} (\bibinfo {year}
  {2022})}\BibitemShut {NoStop}%
\bibitem [{\citenamefont {Nakaoka}\ \emph {et~al.}(2006)\citenamefont
  {Nakaoka}, \citenamefont {Clark}, \citenamefont {Krenner}, \citenamefont
  {Sabathil}, \citenamefont {Bichler}, \citenamefont {Arakawa}, \citenamefont
  {Abstreiter},\ and\ \citenamefont
  {Finley}}]{nakaokaDirectObservationAcoustic2006}%
  \BibitemOpen
  \bibfield  {author} {\bibinfo {author} {\bibfnamefont {T.}~\bibnamefont
  {Nakaoka}}, \bibinfo {author} {\bibfnamefont {E.~C.}\ \bibnamefont {Clark}},
  \bibinfo {author} {\bibfnamefont {H.~J.}\ \bibnamefont {Krenner}}, \bibinfo
  {author} {\bibfnamefont {M.}~\bibnamefont {Sabathil}}, \bibinfo {author}
  {\bibfnamefont {M.}~\bibnamefont {Bichler}}, \bibinfo {author} {\bibfnamefont
  {Y.}~\bibnamefont {Arakawa}}, \bibinfo {author} {\bibfnamefont
  {G.}~\bibnamefont {Abstreiter}},\ and\ \bibinfo {author} {\bibfnamefont
  {J.~J.}\ \bibnamefont {Finley}},\ }\bibfield  {title} {\bibinfo {title}
  {Direct observation of acoustic phonon mediated relaxation between coupled
  exciton states in a single quantum dot molecule},\ }\href
  {https://doi.org/10.1103/PhysRevB.74.121305} {\bibfield  {journal} {\bibinfo
  {journal} {Phys. Rev. B}\ }\textbf {\bibinfo {volume} {74}},\ \bibinfo
  {pages} {121305(R)} (\bibinfo {year} {2006})}\BibitemShut {NoStop}%
\bibitem [{\citenamefont {Borri}\ \emph {et~al.}(2001)\citenamefont {Borri},
  \citenamefont {Langbein}, \citenamefont {Schneider}, \citenamefont {Woggon},
  \citenamefont {Sellin}, \citenamefont {Ouyang},\ and\ \citenamefont
  {Bimberg}}]{Borri2001}%
  \BibitemOpen
  \bibfield  {author} {\bibinfo {author} {\bibfnamefont {P.}~\bibnamefont
  {Borri}}, \bibinfo {author} {\bibfnamefont {W.}~\bibnamefont {Langbein}},
  \bibinfo {author} {\bibfnamefont {S.}~\bibnamefont {Schneider}}, \bibinfo
  {author} {\bibfnamefont {U.}~\bibnamefont {Woggon}}, \bibinfo {author}
  {\bibfnamefont {R.~L.}\ \bibnamefont {Sellin}}, \bibinfo {author}
  {\bibfnamefont {D.}~\bibnamefont {Ouyang}},\ and\ \bibinfo {author}
  {\bibfnamefont {D.}~\bibnamefont {Bimberg}},\ }\bibfield  {title} {\bibinfo
  {title} {{Ultralong Dephasing Time in InGaAs Quantum Dots}},\ }\href
  {https://doi.org/10.1103/PhysRevLett.87.157401} {\bibfield  {journal}
  {\bibinfo  {journal} {Phys. Rev. Lett.}\ }\textbf {\bibinfo {volume} {87}},\
  \bibinfo {pages} {157401} (\bibinfo {year} {2001})}\BibitemShut {NoStop}%
\bibitem [{\citenamefont {Kawa}\ \emph {et~al.}(2022)\citenamefont {Kawa},
  \citenamefont {Kuhn},\ and\ \citenamefont {Machnikowski}}]{kawa2022}%
  \BibitemOpen
  \bibfield  {author} {\bibinfo {author} {\bibfnamefont {K.}~\bibnamefont
  {Kawa}}, \bibinfo {author} {\bibfnamefont {T.}~\bibnamefont {Kuhn}},\ and\
  \bibinfo {author} {\bibfnamefont {P.}~\bibnamefont {Machnikowski}},\
  }\bibfield  {title} {\bibinfo {title} {Coherence limitations in the optical
  control of the singlet-triplet qubit in a quantum dot molecule},\ }\href
  {https://doi.org/10.1103/PhysRevB.106.125308} {\bibfield  {journal} {\bibinfo
   {journal} {Phys. Rev. B}\ }\textbf {\bibinfo {volume} {106}},\ \bibinfo
  {pages} {125308} (\bibinfo {year} {2022})}\BibitemShut {NoStop}%
\bibitem [{\citenamefont {Wiercinski}\ \emph {et~al.}(2023)\citenamefont
  {Wiercinski}, \citenamefont {Gauger},\ and\ \citenamefont
  {Cygorek}}]{Wiercinski2023}%
  \BibitemOpen
  \bibfield  {author} {\bibinfo {author} {\bibfnamefont {J.}~\bibnamefont
  {Wiercinski}}, \bibinfo {author} {\bibfnamefont {E.~M.}\ \bibnamefont
  {Gauger}},\ and\ \bibinfo {author} {\bibfnamefont {M.}~\bibnamefont
  {Cygorek}},\ }\bibfield  {title} {\bibinfo {title} {Phonon coupling versus
  pure dephasing in the photon statistics of cooperative emitters},\ }\href
  {https://doi.org/10.1103/PhysRevResearch.5.013176} {\bibfield  {journal}
  {\bibinfo  {journal} {Phys. Rev. Res.}\ }\textbf {\bibinfo {volume} {5}},\
  \bibinfo {pages} {013176} (\bibinfo {year} {2023})}\BibitemShut {NoStop}%
\bibitem [{\citenamefont {Ramsay}\ \emph {et~al.}(2010)\citenamefont {Ramsay},
  \citenamefont {Gopal}, \citenamefont {Gauger}, \citenamefont {Nazir},
  \citenamefont {Lovett}, \citenamefont {Fox},\ and\ \citenamefont
  {Skolnick}}]{Ramsay2010}%
  \BibitemOpen
  \bibfield  {author} {\bibinfo {author} {\bibfnamefont {A.~J.}\ \bibnamefont
  {Ramsay}}, \bibinfo {author} {\bibfnamefont {A.~V.}\ \bibnamefont {Gopal}},
  \bibinfo {author} {\bibfnamefont {E.~M.}\ \bibnamefont {Gauger}}, \bibinfo
  {author} {\bibfnamefont {A.}~\bibnamefont {Nazir}}, \bibinfo {author}
  {\bibfnamefont {B.~W.}\ \bibnamefont {Lovett}}, \bibinfo {author}
  {\bibfnamefont {A.~M.}\ \bibnamefont {Fox}},\ and\ \bibinfo {author}
  {\bibfnamefont {M.~S.}\ \bibnamefont {Skolnick}},\ }\bibfield  {title}
  {\bibinfo {title} {Damping of exciton rabi rotations by acoustic phonons in
  optically excited $\mathrm{InGaAs}/\mathrm{GaAs}$ quantum dots},\ }\href
  {https://doi.org/10.1103/PhysRevLett.104.017402} {\bibfield  {journal}
  {\bibinfo  {journal} {Phys. Rev. Lett.}\ }\textbf {\bibinfo {volume} {104}},\
  \bibinfo {pages} {017402} (\bibinfo {year} {2010})}\BibitemShut {NoStop}%
\bibitem [{\citenamefont {Vagov}\ \emph {et~al.}(2004)\citenamefont {Vagov},
  \citenamefont {Axt}, \citenamefont {Kuhn}, \citenamefont {Langbein},
  \citenamefont {Borri},\ and\ \citenamefont {Woggon}}]{Vagov2004}%
  \BibitemOpen
  \bibfield  {author} {\bibinfo {author} {\bibfnamefont {A.}~\bibnamefont
  {Vagov}}, \bibinfo {author} {\bibfnamefont {V.~M.}\ \bibnamefont {Axt}},
  \bibinfo {author} {\bibfnamefont {T.}~\bibnamefont {Kuhn}}, \bibinfo {author}
  {\bibfnamefont {W.}~\bibnamefont {Langbein}}, \bibinfo {author}
  {\bibfnamefont {P.}~\bibnamefont {Borri}},\ and\ \bibinfo {author}
  {\bibfnamefont {U.}~\bibnamefont {Woggon}},\ }\bibfield  {title} {\bibinfo
  {title} {Nonmonotonous temperature dependence of the initial decoherence in
  quantum dots},\ }\href {https://doi.org/10.1103/PhysRevB.70.201305}
  {\bibfield  {journal} {\bibinfo  {journal} {Phys. Rev. B}\ }\textbf {\bibinfo
  {volume} {70}},\ \bibinfo {pages} {201305} (\bibinfo {year}
  {2004})}\BibitemShut {NoStop}%
\bibitem [{\citenamefont {Butscher}\ and\ \citenamefont
  {Knorr}(2006)}]{Knorr2006}%
  \BibitemOpen
  \bibfield  {author} {\bibinfo {author} {\bibfnamefont {S.}~\bibnamefont
  {Butscher}}\ and\ \bibinfo {author} {\bibfnamefont {A.}~\bibnamefont
  {Knorr}},\ }\bibfield  {title} {\bibinfo {title} {Theory of strong
  electron–phonon coupling for ultrafast intersubband excitations},\ }\href
  {https://doi.org/https://doi.org/10.1002/pssb.200668078} {\bibfield
  {journal} {\bibinfo  {journal} {physica status solidi (b)}\ }\textbf
  {\bibinfo {volume} {243}},\ \bibinfo {pages} {2423} (\bibinfo {year}
  {2006})}\BibitemShut {NoStop}%
\bibitem [{\citenamefont {Wijesundara}\ \emph {et~al.}(2011)\citenamefont
  {Wijesundara}, \citenamefont {Rolon}, \citenamefont {Ulloa}, \citenamefont
  {Bracker}, \citenamefont {Gammon},\ and\ \citenamefont
  {Stinaff}}]{Wijesundara2011}%
  \BibitemOpen
  \bibfield  {author} {\bibinfo {author} {\bibfnamefont {K.~C.}\ \bibnamefont
  {Wijesundara}}, \bibinfo {author} {\bibfnamefont {J.~E.}\ \bibnamefont
  {Rolon}}, \bibinfo {author} {\bibfnamefont {S.~E.}\ \bibnamefont {Ulloa}},
  \bibinfo {author} {\bibfnamefont {A.~S.}\ \bibnamefont {Bracker}}, \bibinfo
  {author} {\bibfnamefont {D.}~\bibnamefont {Gammon}},\ and\ \bibinfo {author}
  {\bibfnamefont {E.~A.}\ \bibnamefont {Stinaff}},\ }\bibfield  {title}
  {\bibinfo {title} {Tunable exciton relaxation in vertically coupled
  semiconductor inas quantum dots},\ }\href
  {https://doi.org/10.1103/PhysRevB.84.081404} {\bibfield  {journal} {\bibinfo
  {journal} {Phys. Rev. B}\ }\textbf {\bibinfo {volume} {84}},\ \bibinfo
  {pages} {081404} (\bibinfo {year} {2011})}\BibitemShut {NoStop}%
\bibitem [{\citenamefont {Gawarecki}(2018)}]{Gawarecki2018a}%
  \BibitemOpen
  \bibfield  {author} {\bibinfo {author} {\bibfnamefont {K.}~\bibnamefont
  {Gawarecki}},\ }\bibfield  {title} {\bibinfo {title} {{Spin-orbit coupling
  and magnetic-field dependence of carrier states in a self-assembled quantum
  dot}},\ }\href {https://doi.org/10.1103/PhysRevB.97.235408} {\bibfield
  {journal} {\bibinfo  {journal} {Phys. Rev. B}\ }\textbf {\bibinfo {volume}
  {97}},\ \bibinfo {pages} {235408} (\bibinfo {year} {2018})}\BibitemShut
  {NoStop}%
\bibitem [{\citenamefont {Climente}\ \emph {et~al.}(2006)\citenamefont
  {Climente}, \citenamefont {Bertoni}, \citenamefont {Goldoni},\ and\
  \citenamefont {Molinari}}]{Climente2006}%
  \BibitemOpen
  \bibfield  {author} {\bibinfo {author} {\bibfnamefont {J.~I.}\ \bibnamefont
  {Climente}}, \bibinfo {author} {\bibfnamefont {A.}~\bibnamefont {Bertoni}},
  \bibinfo {author} {\bibfnamefont {G.}~\bibnamefont {Goldoni}},\ and\ \bibinfo
  {author} {\bibfnamefont {E.}~\bibnamefont {Molinari}},\ }\bibfield  {title}
  {\bibinfo {title} {Phonon-induced electron relaxation in weakly confined
  single and coupled quantum dots},\ }\href
  {https://doi.org/10.1103/PhysRevB.74.035313} {\bibfield  {journal} {\bibinfo
  {journal} {Phys. Rev. B}\ }\textbf {\bibinfo {volume} {74}},\ \bibinfo
  {pages} {035313} (\bibinfo {year} {2006})}\BibitemShut {NoStop}%
\bibitem [{\citenamefont {Gawarecki}\ \emph {et~al.}(2010)\citenamefont
  {Gawarecki}, \citenamefont {Pochwa\l{}a}, \citenamefont {Grodecka-Grad},\
  and\ \citenamefont {Machnikowski}}]{Gawarecki2010}%
  \BibitemOpen
  \bibfield  {author} {\bibinfo {author} {\bibfnamefont {K.}~\bibnamefont
  {Gawarecki}}, \bibinfo {author} {\bibfnamefont {M.}~\bibnamefont
  {Pochwa\l{}a}}, \bibinfo {author} {\bibfnamefont {A.}~\bibnamefont
  {Grodecka-Grad}},\ and\ \bibinfo {author} {\bibfnamefont {P.}~\bibnamefont
  {Machnikowski}},\ }\bibfield  {title} {\bibinfo {title} {Phonon-assisted
  relaxation and tunneling in self-assembled quantum dot molecules},\ }\href
  {https://doi.org/10.1103/PhysRevB.81.245312} {\bibfield  {journal} {\bibinfo
  {journal} {Phys. Rev. B}\ }\textbf {\bibinfo {volume} {81}},\ \bibinfo
  {pages} {245312} (\bibinfo {year} {2010})}\BibitemShut {NoStop}%
\bibitem [{\citenamefont {Gawarecki}\ and\ \citenamefont
  {Machnikowski}(2012)}]{Gawarecki2012}%
  \BibitemOpen
  \bibfield  {author} {\bibinfo {author} {\bibfnamefont {K.}~\bibnamefont
  {Gawarecki}}\ and\ \bibinfo {author} {\bibfnamefont {P.}~\bibnamefont
  {Machnikowski}},\ }\bibfield  {title} {\bibinfo {title} {Phonon-assisted
  relaxation between hole states in quantum dot molecules},\ }\href
  {https://doi.org/10.1103/PhysRevB.85.041305} {\bibfield  {journal} {\bibinfo
  {journal} {Phys. Rev. B}\ }\textbf {\bibinfo {volume} {85}},\ \bibinfo
  {pages} {041305} (\bibinfo {year} {2012})}\BibitemShut {NoStop}%
\bibitem [{\citenamefont {Grodecka-Grad}\ and\ \citenamefont
  {F\"orstner}(2010)}]{Grodecka-Grad2010}%
  \BibitemOpen
  \bibfield  {author} {\bibinfo {author} {\bibfnamefont {A.}~\bibnamefont
  {Grodecka-Grad}}\ and\ \bibinfo {author} {\bibfnamefont {J.}~\bibnamefont
  {F\"orstner}},\ }\bibfield  {title} {\bibinfo {title} {Theory of
  phonon-mediated relaxation in doped quantum dot molecules},\ }\href
  {https://doi.org/10.1103/PhysRevB.81.115305} {\bibfield  {journal} {\bibinfo
  {journal} {Phys. Rev. B}\ }\textbf {\bibinfo {volume} {81}},\ \bibinfo
  {pages} {115305} (\bibinfo {year} {2010})}\BibitemShut {NoStop}%
\bibitem [{\citenamefont {M\"uller}\ \emph {et~al.}(2012)\citenamefont
  {M\"uller}, \citenamefont {Bechtold}, \citenamefont {Ruppert}, \citenamefont
  {Zecherle}, \citenamefont {Reithmaier}, \citenamefont {Bichler},
  \citenamefont {Krenner}, \citenamefont {Abstreiter}, \citenamefont
  {Holleitner}, \citenamefont {Villas-Boas}, \citenamefont {Betz},\ and\
  \citenamefont {Finley}}]{Mueller2012}%
  \BibitemOpen
  \bibfield  {author} {\bibinfo {author} {\bibfnamefont {K.}~\bibnamefont
  {M\"uller}}, \bibinfo {author} {\bibfnamefont {A.}~\bibnamefont {Bechtold}},
  \bibinfo {author} {\bibfnamefont {C.}~\bibnamefont {Ruppert}}, \bibinfo
  {author} {\bibfnamefont {M.}~\bibnamefont {Zecherle}}, \bibinfo {author}
  {\bibfnamefont {G.}~\bibnamefont {Reithmaier}}, \bibinfo {author}
  {\bibfnamefont {M.}~\bibnamefont {Bichler}}, \bibinfo {author} {\bibfnamefont
  {H.~J.}\ \bibnamefont {Krenner}}, \bibinfo {author} {\bibfnamefont
  {G.}~\bibnamefont {Abstreiter}}, \bibinfo {author} {\bibfnamefont {A.~W.}\
  \bibnamefont {Holleitner}}, \bibinfo {author} {\bibfnamefont {J.~M.}\
  \bibnamefont {Villas-Boas}}, \bibinfo {author} {\bibfnamefont
  {M.}~\bibnamefont {Betz}},\ and\ \bibinfo {author} {\bibfnamefont {J.~J.}\
  \bibnamefont {Finley}},\ }\bibfield  {title} {\bibinfo {title} {Electrical
  control of interdot electron tunneling in a double ingaas quantum-dot
  nanostructure},\ }\href {https://doi.org/10.1103/PhysRevLett.108.197402}
  {\bibfield  {journal} {\bibinfo  {journal} {Phys. Rev. Lett.}\ }\textbf
  {\bibinfo {volume} {108}},\ \bibinfo {pages} {197402} (\bibinfo {year}
  {2012})}\BibitemShut {NoStop}%
\bibitem [{\citenamefont {Krenner}\ \emph {et~al.}(2005)\citenamefont
  {Krenner}, \citenamefont {Sabathil}, \citenamefont {Clark}, \citenamefont
  {Kress}, \citenamefont {Schuh}, \citenamefont {Bichler}, \citenamefont
  {Abstreiter},\ and\ \citenamefont {Finley}}]{Krenner2005}%
  \BibitemOpen
  \bibfield  {author} {\bibinfo {author} {\bibfnamefont {H.~J.}\ \bibnamefont
  {Krenner}}, \bibinfo {author} {\bibfnamefont {M.}~\bibnamefont {Sabathil}},
  \bibinfo {author} {\bibfnamefont {E.~C.}\ \bibnamefont {Clark}}, \bibinfo
  {author} {\bibfnamefont {A.}~\bibnamefont {Kress}}, \bibinfo {author}
  {\bibfnamefont {D.}~\bibnamefont {Schuh}}, \bibinfo {author} {\bibfnamefont
  {M.}~\bibnamefont {Bichler}}, \bibinfo {author} {\bibfnamefont
  {G.}~\bibnamefont {Abstreiter}},\ and\ \bibinfo {author} {\bibfnamefont
  {J.~J.}\ \bibnamefont {Finley}},\ }\bibfield  {title} {\bibinfo {title}
  {Direct observation of controlled coupling in an individual quantum dot
  molecule},\ }\href {https://doi.org/10.1103/PhysRevLett.94.057402} {\bibfield
   {journal} {\bibinfo  {journal} {Phys. Rev. Lett.}\ }\textbf {\bibinfo
  {volume} {94}},\ \bibinfo {pages} {057402} (\bibinfo {year}
  {2005})}\BibitemShut {NoStop}%
\bibitem [{\citenamefont {Bopp}\ \emph
  {et~al.}(2023{\natexlab{b}})\citenamefont {Bopp}, \citenamefont {Cullip},
  \citenamefont {Thalacker}, \citenamefont {Lienhart}, \citenamefont {Schall},
  \citenamefont {Bart}, \citenamefont {Sbresny}, \citenamefont {Boos},
  \citenamefont {Rodt}, \citenamefont {Reuter}, \citenamefont {Ludwig},
  \citenamefont {Wieck}, \citenamefont {Reitzenstein}, \citenamefont {Troiani},
  \citenamefont {Goldoni}, \citenamefont {Molinari}, \citenamefont {M\"uller},\
  and\ \citenamefont {Finley}}]{BoppMagTun}%
  \BibitemOpen
  \bibfield  {author} {\bibinfo {author} {\bibfnamefont {F.}~\bibnamefont
  {Bopp}}, \bibinfo {author} {\bibfnamefont {C.}~\bibnamefont {Cullip}},
  \bibinfo {author} {\bibfnamefont {C.}~\bibnamefont {Thalacker}}, \bibinfo
  {author} {\bibfnamefont {M.}~\bibnamefont {Lienhart}}, \bibinfo {author}
  {\bibfnamefont {J.}~\bibnamefont {Schall}}, \bibinfo {author} {\bibfnamefont
  {N.}~\bibnamefont {Bart}}, \bibinfo {author} {\bibfnamefont {F.}~\bibnamefont
  {Sbresny}}, \bibinfo {author} {\bibfnamefont {K.}~\bibnamefont {Boos}},
  \bibinfo {author} {\bibfnamefont {S.}~\bibnamefont {Rodt}}, \bibinfo {author}
  {\bibfnamefont {D.}~\bibnamefont {Reuter}}, \bibinfo {author} {\bibfnamefont
  {A.}~\bibnamefont {Ludwig}}, \bibinfo {author} {\bibfnamefont {A.~D.}\
  \bibnamefont {Wieck}}, \bibinfo {author} {\bibfnamefont {S.}~\bibnamefont
  {Reitzenstein}}, \bibinfo {author} {\bibfnamefont {F.}~\bibnamefont
  {Troiani}}, \bibinfo {author} {\bibfnamefont {G.}~\bibnamefont {Goldoni}},
  \bibinfo {author} {\bibfnamefont {E.}~\bibnamefont {Molinari}}, \bibinfo
  {author} {\bibfnamefont {K.}~\bibnamefont {M\"uller}},\ and\ \bibinfo
  {author} {\bibfnamefont {J.~J.}\ \bibnamefont {Finley}},\ }\bibfield  {title}
  {\bibinfo {title} {Magnetic tuning of the tunnel coupling in an optically
  active quantum dot molecule},\ }\bibfield  {journal} {\bibinfo  {journal}
  {arXiv:2303.12552v1}\ }\href {https://doi.org/10.48550/arXiv.2303.12552}
  {10.48550/arXiv.2303.12552} (\bibinfo {year}
  {2023}{\natexlab{b}})\BibitemShut {NoStop}%
\bibitem [{\citenamefont {M\"uller}\ \emph {et~al.}(2011)\citenamefont
  {M\"uller}, \citenamefont {Reithmaier}, \citenamefont {Clark}, \citenamefont
  {Jovanov}, \citenamefont {Bichler}, \citenamefont {Krenner}, \citenamefont
  {Betz}, \citenamefont {Abstreiter},\ and\ \citenamefont
  {Finley}}]{Mueller2011}%
  \BibitemOpen
  \bibfield  {author} {\bibinfo {author} {\bibfnamefont {K.}~\bibnamefont
  {M\"uller}}, \bibinfo {author} {\bibfnamefont {G.}~\bibnamefont
  {Reithmaier}}, \bibinfo {author} {\bibfnamefont {E.~C.}\ \bibnamefont
  {Clark}}, \bibinfo {author} {\bibfnamefont {V.}~\bibnamefont {Jovanov}},
  \bibinfo {author} {\bibfnamefont {M.}~\bibnamefont {Bichler}}, \bibinfo
  {author} {\bibfnamefont {H.~J.}\ \bibnamefont {Krenner}}, \bibinfo {author}
  {\bibfnamefont {M.}~\bibnamefont {Betz}}, \bibinfo {author} {\bibfnamefont
  {G.}~\bibnamefont {Abstreiter}},\ and\ \bibinfo {author} {\bibfnamefont
  {J.~J.}\ \bibnamefont {Finley}},\ }\bibfield  {title} {\bibinfo {title}
  {Excited state quantum couplings and optical switching of an artificial
  molecule},\ }\href {https://doi.org/10.1103/PhysRevB.84.081302} {\bibfield
  {journal} {\bibinfo  {journal} {Phys. Rev. B}\ }\textbf {\bibinfo {volume}
  {84}},\ \bibinfo {pages} {081302} (\bibinfo {year} {2011})}\BibitemShut
  {NoStop}%
\bibitem [{\citenamefont {Gawarecki}\ \emph {et~al.}(2023)\citenamefont
  {Gawarecki}, \citenamefont {Spinnler}, \citenamefont {Zhai}, \citenamefont
  {Nguyen}, \citenamefont {Ludwig}, \citenamefont {Warburton}, \citenamefont
  {L\"obl}, \citenamefont {Reiter},\ and\ \citenamefont
  {Machnikowski}}]{Gawarecki2023}%
  \BibitemOpen
  \bibfield  {author} {\bibinfo {author} {\bibfnamefont {K.}~\bibnamefont
  {Gawarecki}}, \bibinfo {author} {\bibfnamefont {C.}~\bibnamefont {Spinnler}},
  \bibinfo {author} {\bibfnamefont {L.}~\bibnamefont {Zhai}}, \bibinfo {author}
  {\bibfnamefont {G.~N.}\ \bibnamefont {Nguyen}}, \bibinfo {author}
  {\bibfnamefont {A.}~\bibnamefont {Ludwig}}, \bibinfo {author} {\bibfnamefont
  {R.~J.}\ \bibnamefont {Warburton}}, \bibinfo {author} {\bibfnamefont {M.~C.}\
  \bibnamefont {L\"obl}}, \bibinfo {author} {\bibfnamefont {D.~E.}\
  \bibnamefont {Reiter}},\ and\ \bibinfo {author} {\bibfnamefont
  {P.}~\bibnamefont {Machnikowski}},\ }\bibfield  {title} {\bibinfo {title}
  {Symmetry breaking via alloy disorder to explain radiative auger transitions
  in self-assembled quantum dots},\ }\href
  {https://doi.org/10.1103/PhysRevB.108.235410} {\bibfield  {journal} {\bibinfo
   {journal} {Phys. Rev. B}\ }\textbf {\bibinfo {volume} {108}},\ \bibinfo
  {pages} {235410} (\bibinfo {year} {2023})}\BibitemShut {NoStop}%
\bibitem [{\citenamefont {L{\"o}bl}\ \emph {et~al.}(2019)\citenamefont
  {L{\"o}bl}, \citenamefont {Scholz}, \citenamefont {S{\"o}llner},
  \citenamefont {Ritzmann}, \citenamefont {Denneulin}, \citenamefont
  {Kov{\'a}cs}, \citenamefont {Kardyna{\l}}, \citenamefont {Wieck},
  \citenamefont {Ludwig},\ and\ \citenamefont {Warburton}}]{Lobl2019a}%
  \BibitemOpen
  \bibfield  {author} {\bibinfo {author} {\bibfnamefont {M.~C.}\ \bibnamefont
  {L{\"o}bl}}, \bibinfo {author} {\bibfnamefont {S.}~\bibnamefont {Scholz}},
  \bibinfo {author} {\bibfnamefont {I.}~\bibnamefont {S{\"o}llner}}, \bibinfo
  {author} {\bibfnamefont {J.}~\bibnamefont {Ritzmann}}, \bibinfo {author}
  {\bibfnamefont {T.}~\bibnamefont {Denneulin}}, \bibinfo {author}
  {\bibfnamefont {A.}~\bibnamefont {Kov{\'a}cs}}, \bibinfo {author}
  {\bibfnamefont {B.~E.}\ \bibnamefont {Kardyna{\l}}}, \bibinfo {author}
  {\bibfnamefont {A.~D.}\ \bibnamefont {Wieck}}, \bibinfo {author}
  {\bibfnamefont {A.}~\bibnamefont {Ludwig}},\ and\ \bibinfo {author}
  {\bibfnamefont {R.~J.}\ \bibnamefont {Warburton}},\ }\bibfield  {title}
  {\bibinfo {title} {Excitons in {{InGaAs}} quantum dots without electron
  wetting layer states},\ }\href {https://doi.org/10.1038/s42005-019-0194-9}
  {\bibfield  {journal} {\bibinfo  {journal} {Commun. Phys.}\ }\textbf
  {\bibinfo {volume} {2}},\ \bibinfo {pages} {93} (\bibinfo {year}
  {2019})}\BibitemShut {NoStop}%
\bibitem [{\citenamefont {Pryor}\ \emph {et~al.}(1998)\citenamefont {Pryor},
  \citenamefont {Kim}, \citenamefont {Wang}, \citenamefont {Williamson},\ and\
  \citenamefont {Zunger}}]{Pryor1998b}%
  \BibitemOpen
  \bibfield  {author} {\bibinfo {author} {\bibfnamefont {C.}~\bibnamefont
  {Pryor}}, \bibinfo {author} {\bibfnamefont {J.}~\bibnamefont {Kim}}, \bibinfo
  {author} {\bibfnamefont {L.~W.}\ \bibnamefont {Wang}}, \bibinfo {author}
  {\bibfnamefont {A.~J.}\ \bibnamefont {Williamson}},\ and\ \bibinfo {author}
  {\bibfnamefont {A.}~\bibnamefont {Zunger}},\ }\bibfield  {title} {\bibinfo
  {title} {{Comparison of two methods for describing the strain profiles in
  quantum dots}},\ }\href@noop {} {\bibfield  {journal} {\bibinfo  {journal}
  {J. Appl. Phys.}\ }\textbf {\bibinfo {volume} {83}},\ \bibinfo {pages} {2548}
  (\bibinfo {year} {1998})}\BibitemShut {NoStop}%
\bibitem [{\citenamefont {Bester}\ \emph {et~al.}(2006)\citenamefont {Bester},
  \citenamefont {Wu}, \citenamefont {Vanderbilt},\ and\ \citenamefont
  {Zunger}}]{Bester2006b}%
  \BibitemOpen
  \bibfield  {author} {\bibinfo {author} {\bibfnamefont {G.}~\bibnamefont
  {Bester}}, \bibinfo {author} {\bibfnamefont {X.}~\bibnamefont {Wu}}, \bibinfo
  {author} {\bibfnamefont {D.}~\bibnamefont {Vanderbilt}},\ and\ \bibinfo
  {author} {\bibfnamefont {A.}~\bibnamefont {Zunger}},\ }\bibfield  {title}
  {\bibinfo {title} {{Importance of Second-Order Piezoelectric Effects in
  Zinc-Blende Semiconductors}},\ }\href
  {https://doi.org/10.1103/PhysRevLett.96.187602} {\bibfield  {journal}
  {\bibinfo  {journal} {Phys. Rev. Lett.}\ }\textbf {\bibinfo {volume} {96}},\
  \bibinfo {pages} {187602} (\bibinfo {year} {2006})}\BibitemShut {NoStop}%
\bibitem [{\citenamefont {Winkler}(2003)}]{Winkler2003}%
  \BibitemOpen
  \bibfield  {author} {\bibinfo {author} {\bibfnamefont {R.}~\bibnamefont
  {Winkler}},\ }\href@noop {} {\emph {\bibinfo {title} {Spin-Orbit Coupling
  Effects in Two-Dimensional Electron and Hole Systems}}}\ (\bibinfo
  {publisher} {Springer},\ \bibinfo {year} {2003})\BibitemShut {NoStop}%
\bibitem [{\citenamefont {Trebin}\ \emph {et~al.}(1979)\citenamefont {Trebin},
  \citenamefont {R{\"{o}}ssler},\ and\ \citenamefont {Ranvaud}}]{Trebin1979}%
  \BibitemOpen
  \bibfield  {author} {\bibinfo {author} {\bibfnamefont {H.~R.}\ \bibnamefont
  {Trebin}}, \bibinfo {author} {\bibfnamefont {U.}~\bibnamefont
  {R{\"{o}}ssler}},\ and\ \bibinfo {author} {\bibfnamefont {R.}~\bibnamefont
  {Ranvaud}},\ }\bibfield  {title} {\bibinfo {title} {{Quantum resonances in
  the valence bands of zinc-blende semiconductors. I. Theoretical aspects}},\
  }\href {https://doi.org/10.1103/PhysRevB.20.686} {\bibfield  {journal}
  {\bibinfo  {journal} {Phys. Rev. B}\ }\textbf {\bibinfo {volume} {20}},\
  \bibinfo {pages} {686} (\bibinfo {year} {1979})}\BibitemShut {NoStop}%
\bibitem [{\citenamefont {Andrzejewski}\ \emph {et~al.}(2010)\citenamefont
  {Andrzejewski}, \citenamefont {Sęk}, \citenamefont {O’Reilly},
  \citenamefont {Fiore},\ and\ \citenamefont {Misiewicz}}]{Andrzejewski2010}%
  \BibitemOpen
  \bibfield  {author} {\bibinfo {author} {\bibfnamefont {J.}~\bibnamefont
  {Andrzejewski}}, \bibinfo {author} {\bibfnamefont {G.}~\bibnamefont {Sęk}},
  \bibinfo {author} {\bibfnamefont {E.}~\bibnamefont {O’Reilly}}, \bibinfo
  {author} {\bibfnamefont {A.}~\bibnamefont {Fiore}},\ and\ \bibinfo {author}
  {\bibfnamefont {J.}~\bibnamefont {Misiewicz}},\ }\bibfield  {title} {\bibinfo
  {title} {{Eight-band k.p calculations of the composition contrast effect on
  the linear polarization properties of columnar quantum dots}},\ }\href
  {https://doi.org/10.1063/1.3346552} {\bibfield  {journal} {\bibinfo
  {journal} {J. Appl. Phys.}\ }\textbf {\bibinfo {volume} {107}},\ \bibinfo
  {pages} {073509} (\bibinfo {year} {2010})}\BibitemShut {NoStop}%
\bibitem [{\citenamefont {Gawe\l{}czyk}\ \emph {et~al.}(2017)\citenamefont
  {Gawe\l{}czyk}, \citenamefont {Syperek}, \citenamefont
  {Mary\ifmmode~\acute{n}\else \'{n}\fi{}ski}, \citenamefont
  {Mrowi\ifmmode~\acute{n}\else \'{n}\fi{}ski}, \citenamefont {Dusanowski},
  \citenamefont {Gawarecki}, \citenamefont {Misiewicz}, \citenamefont {Somers},
  \citenamefont {Reithmaier}, \citenamefont {H\"ofling},\ and\ \citenamefont
  {S\ifmmode~\mbox{\k{e}}\else \k{e}\fi{}k}}]{Gawelczyk2017}%
  \BibitemOpen
  \bibfield  {author} {\bibinfo {author} {\bibfnamefont {M.}~\bibnamefont
  {Gawe\l{}czyk}}, \bibinfo {author} {\bibfnamefont {M.}~\bibnamefont
  {Syperek}}, \bibinfo {author} {\bibfnamefont {A.}~\bibnamefont
  {Mary\ifmmode~\acute{n}\else \'{n}\fi{}ski}}, \bibinfo {author}
  {\bibfnamefont {P.}~\bibnamefont {Mrowi\ifmmode~\acute{n}\else
  \'{n}\fi{}ski}}, \bibinfo {author} {\bibfnamefont {L.}~\bibnamefont
  {Dusanowski}}, \bibinfo {author} {\bibfnamefont {K.}~\bibnamefont
  {Gawarecki}}, \bibinfo {author} {\bibfnamefont {J.}~\bibnamefont
  {Misiewicz}}, \bibinfo {author} {\bibfnamefont {A.}~\bibnamefont {Somers}},
  \bibinfo {author} {\bibfnamefont {J.~P.}\ \bibnamefont {Reithmaier}},
  \bibinfo {author} {\bibfnamefont {S.}~\bibnamefont {H\"ofling}},\ and\
  \bibinfo {author} {\bibfnamefont {G.}~\bibnamefont
  {S\ifmmode~\mbox{\k{e}}\else \k{e}\fi{}k}},\ }\bibfield  {title} {\bibinfo
  {title} {Exciton lifetime and emission polarization dispersion in strongly
  in-plane asymmetric nanostructures},\ }\href
  {https://doi.org/10.1103/PhysRevB.96.245425} {\bibfield  {journal} {\bibinfo
  {journal} {Phys. Rev. B}\ }\textbf {\bibinfo {volume} {96}},\ \bibinfo
  {pages} {245425} (\bibinfo {year} {2017})}\BibitemShut {NoStop}%
\bibitem [{\citenamefont {Feynman}(1939)}]{Feynman1939}%
  \BibitemOpen
  \bibfield  {author} {\bibinfo {author} {\bibfnamefont {R.~P.}\ \bibnamefont
  {Feynman}},\ }\bibfield  {title} {\bibinfo {title} {{Forces in Molecules}},\
  }\href {https://doi.org/10.1103/PhysRev.56.340} {\bibfield  {journal}
  {\bibinfo  {journal} {Phys. Rev.}\ }\textbf {\bibinfo {volume} {56}},\
  \bibinfo {pages} {340} (\bibinfo {year} {1939})}\BibitemShut {NoStop}%
\bibitem [{\citenamefont {{Lew Yan Voon}}\ and\ \citenamefont
  {Ram-Mohan}(1993)}]{LewYanVoon1993}%
  \BibitemOpen
  \bibfield  {author} {\bibinfo {author} {\bibfnamefont {L.~C.}\ \bibnamefont
  {{Lew Yan Voon}}}\ and\ \bibinfo {author} {\bibfnamefont {L.~R.}\
  \bibnamefont {Ram-Mohan}},\ }\bibfield  {title} {\bibinfo {title}
  {{Tight-binding representation of the optical matrix elements: Theory and
  applications}},\ }\href {https://doi.org/10.1103/PhysRevB.47.15500}
  {\bibfield  {journal} {\bibinfo  {journal} {Phys. Rev. B}\ }\textbf {\bibinfo
  {volume} {47}},\ \bibinfo {pages} {15500} (\bibinfo {year}
  {1993})}\BibitemShut {NoStop}%
\bibitem [{\citenamefont {Eissfeller}(2012)}]{Eissfeller2012}%
  \BibitemOpen
  \bibfield  {author} {\bibinfo {author} {\bibfnamefont {T.}~\bibnamefont
  {Eissfeller}},\ }\emph {\bibinfo {title} {Theory of the Electronic Structure
  of Quantum Dots in External Fields}},\ \href@noop {} {Ph.D. thesis},\
  \bibinfo  {school} {Technical University of Munich} (\bibinfo {year}
  {2012})\BibitemShut {NoStop}%
\bibitem [{\citenamefont {Krzykowski}\ \emph {et~al.}(2020)\citenamefont
  {Krzykowski}, \citenamefont {Gawarecki},\ and\ \citenamefont
  {Machnikowski}}]{Krzykowski2020}%
  \BibitemOpen
  \bibfield  {author} {\bibinfo {author} {\bibfnamefont {M.}~\bibnamefont
  {Krzykowski}}, \bibinfo {author} {\bibfnamefont {K.}~\bibnamefont
  {Gawarecki}},\ and\ \bibinfo {author} {\bibfnamefont {P.}~\bibnamefont
  {Machnikowski}},\ }\bibfield  {title} {\bibinfo {title} {Hole spin-flip
  transitions in a self-assembled quantum dot},\ }\href
  {https://doi.org/10.1103/PhysRevB.102.205301} {\bibfield  {journal} {\bibinfo
   {journal} {Phys. Rev. B}\ }\textbf {\bibinfo {volume} {102}},\ \bibinfo
  {pages} {205301} (\bibinfo {year} {2020})}\BibitemShut {NoStop}%
\bibitem [{\citenamefont {Woods}\ \emph {et~al.}(2004)\citenamefont {Woods},
  \citenamefont {Reinecke},\ and\ \citenamefont {Kotlyar}}]{Woods2004}%
  \BibitemOpen
  \bibfield  {author} {\bibinfo {author} {\bibfnamefont {L.~M.}\ \bibnamefont
  {Woods}}, \bibinfo {author} {\bibfnamefont {T.~L.}\ \bibnamefont
  {Reinecke}},\ and\ \bibinfo {author} {\bibfnamefont {R.}~\bibnamefont
  {Kotlyar}},\ }\bibfield  {title} {\bibinfo {title} {Hole spin relaxation in
  quantum dots},\ }\href {https://doi.org/10.1103/PhysRevB.69.125330}
  {\bibfield  {journal} {\bibinfo  {journal} {Phys. Rev. B}\ }\textbf {\bibinfo
  {volume} {69}},\ \bibinfo {pages} {125330} (\bibinfo {year}
  {2004})}\BibitemShut {NoStop}%
\bibitem [{\citenamefont {Wu}\ and\ \citenamefont {Wang}(2014)}]{Wu2014}%
  \BibitemOpen
  \bibinfo {editor} {\bibfnamefont {J.}~\bibnamefont {Wu}}\ and\ \bibinfo
  {editor} {\bibfnamefont {Z.~M.}\ \bibnamefont {Wang}},\ eds.,\ \href@noop {}
  {\emph {\bibinfo {title} {Quantum Dot Molecules}}},\ \bibinfo {series}
  {Lecture {{Notes}} in {{Nanoscale Science}} and {{Technology}}}\ No.\
  \bibinfo {number} {volume 14}\ (\bibinfo  {publisher} {Springer},\ \bibinfo
  {address} {New York},\ \bibinfo {year} {2014})\BibitemShut {NoStop}%
\bibitem [{\citenamefont {Stock}\ \emph {et~al.}(2011)\citenamefont {Stock},
  \citenamefont {Dachner}, \citenamefont {Warming}, \citenamefont {Schliwa},
  \citenamefont {Lochmann}, \citenamefont {Hoffmann}, \citenamefont {Toropov},
  \citenamefont {Bakarov}, \citenamefont {Derebezov}, \citenamefont {Richter},
  \citenamefont {Haisler}, \citenamefont {Knorr},\ and\ \citenamefont
  {Bimberg}}]{Stock2011a}%
  \BibitemOpen
  \bibfield  {author} {\bibinfo {author} {\bibfnamefont {E.}~\bibnamefont
  {Stock}}, \bibinfo {author} {\bibfnamefont {M.-R.}\ \bibnamefont {Dachner}},
  \bibinfo {author} {\bibfnamefont {T.}~\bibnamefont {Warming}}, \bibinfo
  {author} {\bibfnamefont {A.}~\bibnamefont {Schliwa}}, \bibinfo {author}
  {\bibfnamefont {A.}~\bibnamefont {Lochmann}}, \bibinfo {author}
  {\bibfnamefont {A.}~\bibnamefont {Hoffmann}}, \bibinfo {author}
  {\bibfnamefont {A.~I.}\ \bibnamefont {Toropov}}, \bibinfo {author}
  {\bibfnamefont {A.~K.}\ \bibnamefont {Bakarov}}, \bibinfo {author}
  {\bibfnamefont {I.~A.}\ \bibnamefont {Derebezov}}, \bibinfo {author}
  {\bibfnamefont {M.}~\bibnamefont {Richter}}, \bibinfo {author} {\bibfnamefont
  {V.~A.}\ \bibnamefont {Haisler}}, \bibinfo {author} {\bibfnamefont
  {A.}~\bibnamefont {Knorr}},\ and\ \bibinfo {author} {\bibfnamefont
  {D.}~\bibnamefont {Bimberg}},\ }\bibfield  {title} {\bibinfo {title}
  {Acoustic and optical phonon scattering in a single {{In}}({{Ga}}){{As}}
  quantum dot},\ }\href {https://doi.org/10.1103/PhysRevB.83.041304} {\bibfield
   {journal} {\bibinfo  {journal} {Phys. Rev. B}\ }\textbf {\bibinfo {volume}
  {83}},\ \bibinfo {pages} {041304} (\bibinfo {year} {2011})}\BibitemShut
  {NoStop}%
\bibitem [{\citenamefont {Wasilewski}\ \emph {et~al.}(1999)\citenamefont
  {Wasilewski}, \citenamefont {Fafard},\ and\ \citenamefont
  {McCaffrey}}]{Wasilewski1990}%
  \BibitemOpen
  \bibfield  {author} {\bibinfo {author} {\bibfnamefont {Z.}~\bibnamefont
  {Wasilewski}}, \bibinfo {author} {\bibfnamefont {S.}~\bibnamefont {Fafard}},\
  and\ \bibinfo {author} {\bibfnamefont {J.}~\bibnamefont {McCaffrey}},\
  }\bibfield  {title} {\bibinfo {title} {Size and shape engineering of
  vertically stacked self-assembled quantum dots},\ }\href
  {https://doi.org/https://doi.org/10.1016/S0022-0248(98)01539-5} {\bibfield
  {journal} {\bibinfo  {journal} {Journal of Crystal Growth}\ }\textbf
  {\bibinfo {volume} {201-202}},\ \bibinfo {pages} {1131} (\bibinfo {year}
  {1999})}\BibitemShut {NoStop}%
\bibitem [{\citenamefont {Bracker}\ \emph {et~al.}(2006)\citenamefont
  {Bracker}, \citenamefont {Scheibner}, \citenamefont {Doty}, \citenamefont
  {Stinaff}, \citenamefont {Ponomarev}, \citenamefont {Kim}, \citenamefont
  {Whitman}, \citenamefont {Reinecke},\ and\ \citenamefont
  {Gammon}}]{Bracker2006}%
  \BibitemOpen
  \bibfield  {author} {\bibinfo {author} {\bibfnamefont {A.~S.}\ \bibnamefont
  {Bracker}}, \bibinfo {author} {\bibfnamefont {M.}~\bibnamefont {Scheibner}},
  \bibinfo {author} {\bibfnamefont {M.~F.}\ \bibnamefont {Doty}}, \bibinfo
  {author} {\bibfnamefont {E.~A.}\ \bibnamefont {Stinaff}}, \bibinfo {author}
  {\bibfnamefont {I.~V.}\ \bibnamefont {Ponomarev}}, \bibinfo {author}
  {\bibfnamefont {J.~C.}\ \bibnamefont {Kim}}, \bibinfo {author} {\bibfnamefont
  {L.~J.}\ \bibnamefont {Whitman}}, \bibinfo {author} {\bibfnamefont {T.~L.}\
  \bibnamefont {Reinecke}},\ and\ \bibinfo {author} {\bibfnamefont
  {D.}~\bibnamefont {Gammon}},\ }\bibfield  {title} {\bibinfo {title}
  {{Engineering electron and hole tunneling with asymmetric InAs quantum dot
  molecules}},\ }\href {https://doi.org/10.1063/1.2400397} {\bibfield
  {journal} {\bibinfo  {journal} {Applied Physics Letters}\ }\textbf {\bibinfo
  {volume} {89}},\ \bibinfo {pages} {233110} (\bibinfo {year}
  {2006})}\BibitemShut {NoStop}%
\bibitem [{\citenamefont {Schall}\ \emph {et~al.}(2021)\citenamefont {Schall},
  \citenamefont {Deconinck}, \citenamefont {Bart}, \citenamefont {Florian},
  \citenamefont {von Helversen}, \citenamefont {Dangel}, \citenamefont
  {Schmidt}, \citenamefont {Bremer}, \citenamefont {Bopp}, \citenamefont
  {Hüllen}, \citenamefont {Gies}, \citenamefont {Reuter}, \citenamefont
  {Wieck}, \citenamefont {Rodt}, \citenamefont {Finley}, \citenamefont
  {Jahnke}, \citenamefont {Ludwig},\ and\ \citenamefont
  {Reitzenstein}}]{Schall2021}%
  \BibitemOpen
  \bibfield  {author} {\bibinfo {author} {\bibfnamefont {J.}~\bibnamefont
  {Schall}}, \bibinfo {author} {\bibfnamefont {M.}~\bibnamefont {Deconinck}},
  \bibinfo {author} {\bibfnamefont {N.}~\bibnamefont {Bart}}, \bibinfo {author}
  {\bibfnamefont {M.}~\bibnamefont {Florian}}, \bibinfo {author} {\bibfnamefont
  {M.}~\bibnamefont {von Helversen}}, \bibinfo {author} {\bibfnamefont
  {C.}~\bibnamefont {Dangel}}, \bibinfo {author} {\bibfnamefont
  {R.}~\bibnamefont {Schmidt}}, \bibinfo {author} {\bibfnamefont
  {L.}~\bibnamefont {Bremer}}, \bibinfo {author} {\bibfnamefont
  {F.}~\bibnamefont {Bopp}}, \bibinfo {author} {\bibfnamefont {I.}~\bibnamefont
  {Hüllen}}, \bibinfo {author} {\bibfnamefont {C.}~\bibnamefont {Gies}},
  \bibinfo {author} {\bibfnamefont {D.}~\bibnamefont {Reuter}}, \bibinfo
  {author} {\bibfnamefont {A.~D.}\ \bibnamefont {Wieck}}, \bibinfo {author}
  {\bibfnamefont {S.}~\bibnamefont {Rodt}}, \bibinfo {author} {\bibfnamefont
  {J.~J.}\ \bibnamefont {Finley}}, \bibinfo {author} {\bibfnamefont
  {F.}~\bibnamefont {Jahnke}}, \bibinfo {author} {\bibfnamefont
  {A.}~\bibnamefont {Ludwig}},\ and\ \bibinfo {author} {\bibfnamefont
  {S.}~\bibnamefont {Reitzenstein}},\ }\bibfield  {title} {\bibinfo {title}
  {Bright electrically controllable quantum-dot-molecule devices fabricated by
  in situ electron-beam lithography},\ }\href
  {https://doi.org/https://doi.org/10.1002/qute.202100002} {\bibfield
  {journal} {\bibinfo  {journal} {Advanced Quantum Technologies}\ }\textbf
  {\bibinfo {volume} {4}},\ \bibinfo {pages} {2100002} (\bibinfo {year}
  {2021})}\BibitemShut {NoStop}%
\bibitem [{\citenamefont {Narvaez}\ \emph {et~al.}(2006)\citenamefont
  {Narvaez}, \citenamefont {Bester},\ and\ \citenamefont
  {Zunger}}]{Narvaez2006}%
  \BibitemOpen
  \bibfield  {author} {\bibinfo {author} {\bibfnamefont {G.~A.}\ \bibnamefont
  {Narvaez}}, \bibinfo {author} {\bibfnamefont {G.}~\bibnamefont {Bester}},\
  and\ \bibinfo {author} {\bibfnamefont {A.}~\bibnamefont {Zunger}},\
  }\bibfield  {title} {\bibinfo {title} {Carrier relaxation mechanisms in
  self-assembled (in,ga)as/gaas quantum dots: Efficient p-->s auger relaxation
  of electrons},\ }\href {https://doi.org/10.1103/PhysRevB.74.075403}
  {\bibfield  {journal} {\bibinfo  {journal} {Phys. Rev. B}\ }\textbf {\bibinfo
  {volume} {74}},\ \bibinfo {pages} {075403} (\bibinfo {year}
  {2006})}\BibitemShut {NoStop}%
\bibitem [{\citenamefont {Li}\ \emph {et~al.}(1999)\citenamefont {Li},
  \citenamefont {Nakayama},\ and\ \citenamefont {Arakawa}}]{Li1999}%
  \BibitemOpen
  \bibfield  {author} {\bibinfo {author} {\bibfnamefont {X.-Q.}\ \bibnamefont
  {Li}}, \bibinfo {author} {\bibfnamefont {H.}~\bibnamefont {Nakayama}},\ and\
  \bibinfo {author} {\bibfnamefont {Y.}~\bibnamefont {Arakawa}},\ }\bibfield
  {title} {\bibinfo {title} {Phonon bottleneck in quantum dots: Role of
  lifetime of the confined optical phonons},\ }\href
  {https://doi.org/10.1103/PhysRevB.59.5069} {\bibfield  {journal} {\bibinfo
  {journal} {Phys. Rev. B}\ }\textbf {\bibinfo {volume} {59}},\ \bibinfo
  {pages} {5069} (\bibinfo {year} {1999})}\BibitemShut {NoStop}%
\bibitem [{\citenamefont {Urayama}\ \emph {et~al.}(2001)\citenamefont
  {Urayama}, \citenamefont {Norris}, \citenamefont {Singh},\ and\ \citenamefont
  {Bhattacharya}}]{Urayama2001}%
  \BibitemOpen
  \bibfield  {author} {\bibinfo {author} {\bibfnamefont {J.}~\bibnamefont
  {Urayama}}, \bibinfo {author} {\bibfnamefont {T.~B.}\ \bibnamefont {Norris}},
  \bibinfo {author} {\bibfnamefont {J.}~\bibnamefont {Singh}},\ and\ \bibinfo
  {author} {\bibfnamefont {P.}~\bibnamefont {Bhattacharya}},\ }\bibfield
  {title} {\bibinfo {title} {Observation of phonon bottleneck in quantum dot
  electronic relaxation},\ }\href {https://doi.org/10.1103/PhysRevLett.86.4930}
  {\bibfield  {journal} {\bibinfo  {journal} {Phys. Rev. Lett.}\ }\textbf
  {\bibinfo {volume} {86}},\ \bibinfo {pages} {4930} (\bibinfo {year}
  {2001})}\BibitemShut {NoStop}%
\bibitem [{\citenamefont {Bacher}\ \emph {et~al.}(1999)\citenamefont {Bacher},
  \citenamefont {Weigand}, \citenamefont {Seufert}, \citenamefont
  {Kulakovskii}, \citenamefont {Gippius}, \citenamefont {Forchel},
  \citenamefont {Leonardi},\ and\ \citenamefont {Hommel}}]{Bacher1999}%
  \BibitemOpen
  \bibfield  {author} {\bibinfo {author} {\bibfnamefont {G.}~\bibnamefont
  {Bacher}}, \bibinfo {author} {\bibfnamefont {R.}~\bibnamefont {Weigand}},
  \bibinfo {author} {\bibfnamefont {J.}~\bibnamefont {Seufert}}, \bibinfo
  {author} {\bibfnamefont {V.~D.}\ \bibnamefont {Kulakovskii}}, \bibinfo
  {author} {\bibfnamefont {N.~A.}\ \bibnamefont {Gippius}}, \bibinfo {author}
  {\bibfnamefont {A.}~\bibnamefont {Forchel}}, \bibinfo {author} {\bibfnamefont
  {K.}~\bibnamefont {Leonardi}},\ and\ \bibinfo {author} {\bibfnamefont
  {D.}~\bibnamefont {Hommel}},\ }\bibfield  {title} {\bibinfo {title}
  {Biexciton versus exciton lifetime in a single semiconductor quantum dot},\
  }\href {https://doi.org/10.1103/PhysRevLett.83.4417} {\bibfield  {journal}
  {\bibinfo  {journal} {Phys. Rev. Lett.}\ }\textbf {\bibinfo {volume} {83}},\
  \bibinfo {pages} {4417} (\bibinfo {year} {1999})}\BibitemShut {NoStop}%
\bibitem [{\citenamefont {{Van Den Zegel}}\ \emph {et~al.}(1986)\citenamefont
  {{Van Den Zegel}}, \citenamefont {Boens}, \citenamefont {Daems},\ and\
  \citenamefont {{De Schryver}}}]{Zegel1986}%
  \BibitemOpen
  \bibfield  {author} {\bibinfo {author} {\bibfnamefont {M.}~\bibnamefont {{Van
  Den Zegel}}}, \bibinfo {author} {\bibfnamefont {N.}~\bibnamefont {Boens}},
  \bibinfo {author} {\bibfnamefont {D.}~\bibnamefont {Daems}},\ and\ \bibinfo
  {author} {\bibfnamefont {F.}~\bibnamefont {{De Schryver}}},\ }\bibfield
  {title} {\bibinfo {title} {Possibilities and limitations of the
  time-correlated single photon counting technique: a comparative study of
  correction methods for the wavelength dependence of the instrument response
  function},\ }\href
  {https://doi.org/https://doi.org/10.1016/0301-0104(86)85096-0} {\bibfield
  {journal} {\bibinfo  {journal} {Chem. Phys.}\ }\textbf {\bibinfo {volume}
  {101}},\ \bibinfo {pages} {311} (\bibinfo {year} {1986})}\BibitemShut
  {NoStop}%
\bibitem [{\citenamefont {Świderski}(2016)}]{Swiderski2016}%
  \BibitemOpen
  \bibfield  {author} {\bibinfo {author} {\bibfnamefont {M.}~\bibnamefont
  {Świderski}, \bibfnamefont {M.~Zieliński}},\ }\bibfield  {title} {\bibinfo
  {title} {Exact diagonalization approach for atomistic calculation of
  piezoelectric effects in semiconductor quantum dots},\ }\href
  {https://doi.org/10.12693/APhysPolA.129.A-79} {\bibfield  {journal} {\bibinfo
   {journal} {Acta Phys. Pol., A}\ }\textbf {\bibinfo {volume} {129}},\
  \bibinfo {pages} {79} (\bibinfo {year} {2016})}\BibitemShut {NoStop}%
\bibitem [{\citenamefont {Thr\"anhardt}\ \emph {et~al.}(2002)\citenamefont
  {Thr\"anhardt}, \citenamefont {Ell}, \citenamefont {Khitrova},\ and\
  \citenamefont {Gibbs}}]{Thranhardt2002}%
  \BibitemOpen
  \bibfield  {author} {\bibinfo {author} {\bibfnamefont {A.}~\bibnamefont
  {Thr\"anhardt}}, \bibinfo {author} {\bibfnamefont {C.}~\bibnamefont {Ell}},
  \bibinfo {author} {\bibfnamefont {G.}~\bibnamefont {Khitrova}},\ and\
  \bibinfo {author} {\bibfnamefont {H.~M.}\ \bibnamefont {Gibbs}},\ }\bibfield
  {title} {\bibinfo {title} {Relation between dipole moment and radiative
  lifetime in interface fluctuation quantum dots},\ }\href
  {https://doi.org/10.1103/PhysRevB.65.035327} {\bibfield  {journal} {\bibinfo
  {journal} {Phys. Rev. B}\ }\textbf {\bibinfo {volume} {65}},\ \bibinfo
  {pages} {035327} (\bibinfo {year} {2002})}\BibitemShut {NoStop}%
\bibitem [{\citenamefont {Amirtharaj}\ and\ \citenamefont
  {Seiler}(1994)}]{Amirtharaj1994}%
  \BibitemOpen
  \bibfield  {author} {\bibinfo {author} {\bibfnamefont {P.}~\bibnamefont
  {Amirtharaj}}\ and\ \bibinfo {author} {\bibfnamefont {D.}~\bibnamefont
  {Seiler}},\ }\href@noop {} {\emph {\bibinfo {title} {Optical Properties of
  Semiconductors, in Handbook of Optics, Vol. 2: Devices, Measurements, and
  Properties, 2nd ed.}}}\ (\bibinfo  {publisher} {McGraw-Hill Professional, New
  York, San Francisco},\ \bibinfo {year} {1994})\BibitemShut {NoStop}%
\bibitem [{\citenamefont {Daniels}\ \emph {et~al.}(2013)\citenamefont
  {Daniels}, \citenamefont {Machnikowski},\ and\ \citenamefont
  {Kuhn}}]{Daniels2013}%
  \BibitemOpen
  \bibfield  {author} {\bibinfo {author} {\bibfnamefont {J.~M.}\ \bibnamefont
  {Daniels}}, \bibinfo {author} {\bibfnamefont {P.}~\bibnamefont
  {Machnikowski}},\ and\ \bibinfo {author} {\bibfnamefont {T.}~\bibnamefont
  {Kuhn}},\ }\bibfield  {title} {\bibinfo {title} {Excitons in quantum dot
  molecules: Coulomb coupling, spin-orbit effects, and phonon-induced line
  broadening},\ }\href {https://doi.org/10.1103/PhysRevB.88.205307} {\bibfield
  {journal} {\bibinfo  {journal} {Phys. Rev. B}\ }\textbf {\bibinfo {volume}
  {88}},\ \bibinfo {pages} {205307} (\bibinfo {year} {2013})}\BibitemShut
  {NoStop}%
\bibitem [{\citenamefont {Virtanen}\ \emph {et~al.}(2020)\citenamefont
  {Virtanen}, \citenamefont {Gommers}, \citenamefont {Oliphant}, \citenamefont
  {Haberland}, \citenamefont {Reddy}, \citenamefont {Cournapeau}, \citenamefont
  {Burovski}, \citenamefont {Peterson}, \citenamefont {Weckesser},
  \citenamefont {Bright}, \citenamefont {{van der Walt}}, \citenamefont
  {Brett}, \citenamefont {Wilson}, \citenamefont {Millman}, \citenamefont
  {Mayorov}, \citenamefont {Nelson}, \citenamefont {Jones}, \citenamefont
  {Kern}, \citenamefont {Larson}, \citenamefont {Carey}, \citenamefont {Polat},
  \citenamefont {Feng}, \citenamefont {Moore}, \citenamefont {{VanderPlas}},
  \citenamefont {Laxalde}, \citenamefont {Perktold}, \citenamefont {Cimrman},
  \citenamefont {Henriksen}, \citenamefont {Quintero}, \citenamefont {Harris},
  \citenamefont {Archibald}, \citenamefont {Ribeiro}, \citenamefont
  {Pedregosa}, \citenamefont {{van Mulbregt}},\ and\ \citenamefont {{SciPy 1.0
  Contributors}}}]{2020SciPy-NMeth}%
  \BibitemOpen
  \bibfield  {author} {\bibinfo {author} {\bibfnamefont {P.}~\bibnamefont
  {Virtanen}}, \bibinfo {author} {\bibfnamefont {R.}~\bibnamefont {Gommers}},
  \bibinfo {author} {\bibfnamefont {T.~E.}\ \bibnamefont {Oliphant}}, \bibinfo
  {author} {\bibfnamefont {M.}~\bibnamefont {Haberland}}, \bibinfo {author}
  {\bibfnamefont {T.}~\bibnamefont {Reddy}}, \bibinfo {author} {\bibfnamefont
  {D.}~\bibnamefont {Cournapeau}}, \bibinfo {author} {\bibfnamefont
  {E.}~\bibnamefont {Burovski}}, \bibinfo {author} {\bibfnamefont
  {P.}~\bibnamefont {Peterson}}, \bibinfo {author} {\bibfnamefont
  {W.}~\bibnamefont {Weckesser}}, \bibinfo {author} {\bibfnamefont
  {J.}~\bibnamefont {Bright}}, \bibinfo {author} {\bibfnamefont {S.~J.}\
  \bibnamefont {{van der Walt}}}, \bibinfo {author} {\bibfnamefont
  {M.}~\bibnamefont {Brett}}, \bibinfo {author} {\bibfnamefont
  {J.}~\bibnamefont {Wilson}}, \bibinfo {author} {\bibfnamefont {K.~J.}\
  \bibnamefont {Millman}}, \bibinfo {author} {\bibfnamefont {N.}~\bibnamefont
  {Mayorov}}, \bibinfo {author} {\bibfnamefont {A.~R.~J.}\ \bibnamefont
  {Nelson}}, \bibinfo {author} {\bibfnamefont {E.}~\bibnamefont {Jones}},
  \bibinfo {author} {\bibfnamefont {R.}~\bibnamefont {Kern}}, \bibinfo {author}
  {\bibfnamefont {E.}~\bibnamefont {Larson}}, \bibinfo {author} {\bibfnamefont
  {C.~J.}\ \bibnamefont {Carey}}, \bibinfo {author} {\bibfnamefont
  {{\.I}.}~\bibnamefont {Polat}}, \bibinfo {author} {\bibfnamefont
  {Y.}~\bibnamefont {Feng}}, \bibinfo {author} {\bibfnamefont {E.~W.}\
  \bibnamefont {Moore}}, \bibinfo {author} {\bibfnamefont {J.}~\bibnamefont
  {{VanderPlas}}}, \bibinfo {author} {\bibfnamefont {D.}~\bibnamefont
  {Laxalde}}, \bibinfo {author} {\bibfnamefont {J.}~\bibnamefont {Perktold}},
  \bibinfo {author} {\bibfnamefont {R.}~\bibnamefont {Cimrman}}, \bibinfo
  {author} {\bibfnamefont {I.}~\bibnamefont {Henriksen}}, \bibinfo {author}
  {\bibfnamefont {E.~A.}\ \bibnamefont {Quintero}}, \bibinfo {author}
  {\bibfnamefont {C.~R.}\ \bibnamefont {Harris}}, \bibinfo {author}
  {\bibfnamefont {A.~M.}\ \bibnamefont {Archibald}}, \bibinfo {author}
  {\bibfnamefont {A.~H.}\ \bibnamefont {Ribeiro}}, \bibinfo {author}
  {\bibfnamefont {F.}~\bibnamefont {Pedregosa}}, \bibinfo {author}
  {\bibfnamefont {P.}~\bibnamefont {{van Mulbregt}}},\ and\ \bibinfo {author}
  {\bibnamefont {{SciPy 1.0 Contributors}}},\ }\bibfield  {title} {\bibinfo
  {title} {{{SciPy} 1.0: Fundamental Algorithms for Scientific Computing in
  Python}},\ }\href {https://doi.org/10.1038/s41592-019-0686-2} {\bibfield
  {journal} {\bibinfo  {journal} {Nat. Methods}\ }\textbf {\bibinfo {volume}
  {17}},\ \bibinfo {pages} {261} (\bibinfo {year} {2020})}\BibitemShut
  {NoStop}%
\end{thebibliography}%

\end{document}